\pdfoutput=1

\documentclass[sigplan,screen,usenames,dvipsnames,9pt]{acmart}
\usepackage{flushend} 

\usepackage{stmaryrd} 

\usepackage{amsmath}
\usepackage{amsthm}

\usepackage{hyperref}

\usepackage{graphicx}
\usepackage{listings}
\usepackage{mathpartir}

\usepackage{paralist} 

\usepackage{cleveref}
\crefname{section}{\S}{\S\S}
\Crefname{section}{\S}{\S\S}
\crefname{figure}{fig.}{figs.}
\Crefname{Figure}{Fig.}{Figs.}

\usepackage{enumitem}
\usepackage{multirow}
\usepackage[]{xcolor}

\usepackage{todonotes}

\newcommand{\rn}[1]{\pgwrapper{RRN}{#1}}  

\InputIfFileExists{activateeditingmarks}{
}{
    \def\noeditingmarks{}
}

\definecolor{comment-red}{rgb}{0.8,0,0}
\definecolor{dark-green}{rgb}{0.0,0.4,0}
\definecolor{dark-blue}{rgb}{0.0,0.0,0.55}
\definecolor{very-dark-green}{rgb}{0.0,0.3,0}

\ifx\noeditingmarks\undefined

   \newcommand{\new}[1]{{\color{very-dark-green}{#1}}}
   \newcommand{\nnew}[1]{{\color{dark-blue}{#1}}}

   \definecolor{mygrey}{rgb}{0.7,0.7,0.7}

   \newcommand{\pgwrapper}[2]{\todo[size=\footnotesize]{#1: #2}}

   \InputIfFileExists{activategreybg}{
       \definecolor{comment-red}{rgb}{0.5,0,0}
       \pagecolor{mygrey}
   }{}

  \newcommand{\rnfloat}[1]{\pgwrapper{RRN}{#1}\vspace{-4mm}}  

\else

   \newcommand{\pgwrapper}[2]{}

   \newcommand{\new}[1]{#1}
   \newcommand{\nnew}[1]{#1}

   \newcommand{\rnfloat}[1]{}  
\fi

\usepackage{listings}
\usepackage{color}
\usepackage{amssymb}
\definecolor{darkgreen}{rgb}{0,0.5,0}
\definecolor{darkred}{rgb}{0.5,0,0}

\newcommand{\il}[1]{\lstinline{#1}}

\definecolor{mygray}{rgb}{0.5,0.5,0.5}

\usepackage{upquote}
\usepackage{listings}
\usepackage{xcolor}

\definecolor{darkgreen}{rgb}{0,0.5,0}
\definecolor{darkred}{rgb}{0.5,0,0}

\lstdefinestyle{haskell}{%
    language=Haskell,
    upquote=true,
    basicstyle=\footnotesize\ttfamily,
    keywordstyle=\bfseries,
    deletekeywords={case,class,data,default,deriving,do,in,instance,let,of,type,where},
    morekeywords={[2]class,data,default,deriving,family,instance,type,where,if,then,else},
    morekeywords={[3]in,let,case,of,do,module,import},
    literate=
        {\\}{{$\lambda$}}1
        {\\\\}{{\char`\\\char`\\}}1
        {>->}{>->}3
        {>>=}{>>=}3
        {->}{{$\rightarrow$}}2
        {>=}{{$\geq$}}2
        {<-}{{$\leftarrow$}}2
        {<=}{{$\leq$}}2
        {=>}{{$\Rightarrow$}}2
        {|}{{$\mid$}}1
        {forall}{{$\forall$}}1
        {exists}{{$\exists$}}1
        {...}{{$\cdots$}}3,
    moredelim=**[is][\color{gray}]{@}{@}
}

\lstdefinestyle{inline}{%
    style=haskell,
    basicstyle=\footnotesize\ttfamily,
    keywordstyle=[1],
    keywordstyle=[2],
    keywordstyle=[3],
    keywordstyle=[4],
    literate=
        {\\}{{$\lambda$}}1
        {\\\\}{{\char`\\\char`\\}}1
        {>->}{>->}3
        {>>=}{>>=}3
        {->}{{$\rightarrow$\space}}3    
        {>=}{{$\geq$}}2
        {<-}{{$\leftarrow$}}2
        {<=}{{$\leq$}}2
        {=>}{{$\Rightarrow$}}2
        {|}{{$\mid$}}1
        {forall}{{$\forall$}}1
        {exists}{{$\exists$}}1
        {...}{{$\cdots$}}3
}

\lstdefinestyle{snippet}{
        style=haskell,
        numbers=none,
        xleftmargin=0.0ex,
        aboveskip=3pt,
        belowskip=3pt
}

\lstnewenvironment{snippet}
    {\lstset{style=snippet,name=snippet}}
    {}

\makeatletter
\lstnewenvironment{code}[1][]
    {\lstset{style=haskell,
    firstnumber=auto,
    name=code,
    #1}%
    \csname\@lst @SetFirstNumber\endcsname}
    {\csname\@lst @SaveFirstNumber\endcsname}
\makeatother

\newcommand{\makeatchar}{\lstDeleteShortInline@}

\newcommand{\mypara}[1]{\vspace{-1.2mm}\paragraph{{#1}:}}

\begin{document}


\copyrightyear{2017} 
\acmYear{2017} 
\setcopyright{acmlicensed}
\acmConference{Haskell'17}{September 7--8, 2017}{Oxford, United Kingdom}
\acmPrice{15.00}
\acmDOI{10.1145/3122955.3122973}
\acmISBN{978-1-4503-5182-9/17/09}

\ccsdesc[500]{Computing methodologies~Concurrent algorithms}
\ccsdesc[300]{Software and its engineering~Functional languages}


\citestyle{acmauthoryear}

\author{Chao-Hong Chen}
\affiliation{
  \institution{Indiana University, USA}            
}
\email{chen464@indiana.edu}          

\author{Vikraman Choudhury}
\affiliation{
  \institution{Indiana University, USA}            
}
\email{vikraman@indiana.edu}          

\author{Ryan R. Newton}
\affiliation{
  \institution{Indiana University, USA}            
}
\email{rrnewton@indiana.edu}          



\title{Adaptive Lock-Free Data Structures in Haskell:\\A General Method for Concurrent Implementation Swapping}

\begin{abstract}

A key part of implementing high-level languages is providing built-in and
default data structures.  Yet selecting good defaults is hard.  A mutable data
structure's workload is not known in advance, and it may shift over its
lifetime---e.g., between read-heavy and write-heavy, or from heavy contention by
multiple threads to single-threaded or low-frequency use.
One idea is to switch implementations adaptively,
but it is nontrivial to switch the implementation of a concurrent data structure
at runtime.  Performing the transition requires a concurrent snapshot of data
structure contents, which normally demands special engineering in the data
structure's design.
\new{However, in this paper we identify and formalize an relevant property of lock-free
  algorithms.  Namely, lock-freedom is sufficient to guarantee that {\em
    freezing} memory locations in an {\em arbitrary} order will result in a valid
  snapshot.}

{Several functional languages have data structures that {freeze} and {thaw},
  transitioning between mutable and immutable, such as Haskell vectors and
  Clojure transients, \new{but these enable only single-threaded writers.}}
We generalize this approach to augment an {arbitrary} lock-free data structure
with the ability to gradually freeze and optionally transition to a new
representation.  This augmentation doesn't require changing the algorithm or
code for the data structure, only replacing its datatype for mutable references
with a freezable variant.
In this paper, we present an algorithm for lifting plain to adaptive data and
prove that the resulting hybrid data structure is itself lock-free,
linearizable, and simulates the original.  We also perform an empirical case
study in the context of {\em heating up} and {\em cooling down} concurrent maps.


\end{abstract}

\keywords{lock-free algorithms, concurrent data structures, concurrency, parallelism}

\maketitle


\lstMakeShortInline[]@







\newcommand{\Pool}{\ensuremath{\mathcal{P}}}
\newcommand{\Trace}{\ensuremath{\mathcal{T}r}}
\newcommand{\seq}[1]{\ensuremath{\langle #1 \rangle}}
\newcommand{\emptyoptrace}{\seq{}}
\newcommand{\seqcons}[2]{\ensuremath{#1 \cdot #2}}

\newcommand{\rulename}[1]{#1}  
\newcommand{\ouremptyset}{\ensuremath{\emptyset}}

\newcommand{\results}[1]{\ensuremath{results(#1)}}
\newcommand{\return}[1]{\ensuremath{return(#1)}}
\newcommand{\methods}[1]{\ensuremath{methods\allowbreak(#1)}}

\newcommand{\uninit}{\ensuremath{uninit}}
\newcommand{\Frzn}{\ensuremath{Frzn}}

\newcommand{\clients}{\ensuremath{\mathbb{C}}}
\newcommand{\threadids}{\ensuremath{\mathfrak{T}}}
\newcommand{\none}{\ensuremath{\textsf{none}}}
\newcommand{\updateC}[3]{\ensuremath{C[#1 \xleftarrow{#2} #3]}}

\newcommand{\stepsto}[1]{\ensuremath{\; \rightarrow_{#1} \; }}

\newcommand\doubleplus{+\kern-1.3ex+\kern0.8ex}


\section{Introduction}\label{sec:intro}

High-level, productivity languages are equipped with rich built-in data
structures, such as mutable and immutable dictionaries.
If we envision parallel-by-default languages of the future,
programmers may expect that built-in data structures support concurrency too.
This will require intelligent selections of {\em default implementations} to not incur
undue overhead under either contended or single-threaded access patterns.

If you need a one-size-fits all choice,
automatically selecting or swapping between data structure
implementations at runtime
has a natural appeal~\cite{java-collection-swapper,pypy}.
These swaps can match implementations to observed workloads: e.g., relative
frequency of methods, degree of contention, or data structure contents.
In the sequential case, {\em adaptive} data structures are already used to good
effect by tracing JIT compilers for dynamic languages such as PyPy and
Pycket~\cite{Bauman:2015:PTJ:2784731.2784740}, which, for instance,
optimistically assumes that an array of floats will continue to contain only
floats, and fall back automatically to a more general representation only if
this is violated.

But such {adaptive} data structures are {\em much} harder to achieve in
concurrent settings than in sequential.  While implementations of adaptive data
structures exist in Java and other
languages~\cite{Kusum:2016:SFA:2892208.2892220,java-collection-swapper,DeWael:2015:JDS:2814228.2814231},
no techniques exist to apply this form of adaptation in a multithreaded
scenario.  How can we convert a data structure that is being mutated
concurrently? {{The problem is similar to that of concurrent garbage collection,
    where a collector tries to move an object graph while the mutator modifies
    it.}}

In this paper we introduce a {general} approach for freezing a concurrent, lock-free,
{\em mutable} data structure---\nnew{subject only to a few restrictions on the
  shape of methods and which instructions linearize}.
 We do this using the simple notion of a {\em freezable
  reference} that replaces the regular mutable reference.
%
Remarkably, it is possible to freeze the references within a data structure {\em
  in an arbitrary order}, and still have a valid snapshot.  The snapshot is valid
because all operations on the original structure are linearizable, which
prevents intermediate modifications due to half-completed methods from corrupting
the structure or affecting its logical contents.

Using freezable references, we show how to build a hybrid data structure that
transitions between two lock-free representations, while in turn preserving
lock-freedom and linearizability for the complete structure.
\new{The end result is a hybrid exposing a special method to trigger transition
  between representations, with user-exposed {\em snapshotting} being a special
  case of transitioning to an immutable data structure inside a reference that
  allows $O(1)$ exact snapshots.}

The contributions of this paper are:
\begin{itemize}[topsep=0pt, noitemsep]
\item \new{We prove a property of lock-free data structures that has not
  previously been formalized: that the memory locations making up the
  structure can be frozen in arbitrary order, providing a valid snapshot state.
  We apply the tools of operational semantics in this concurrency proof, which
  enables more precision than is standard for such proofs.}

\item We present a novel algorithm for building hybrid, adaptive data structures
  which we prove is lock-free, linearizable, and correctly models the original
  structures (\Cref{sec:correctness}).  Our formal model uses an abstract
  \new{machine that models clients interacting with a lock free data structure}.

\item We demonstrate how to apply the method in a Haskell library that
  transitions between a concurrent Ctrie~\cite{Prokopec:2012:CTE:2145816.2145836}
  and a purely functional hashmap (\Cref{sec:app}).
  We perform a case study measuring the benefit of transitioning from a
  concurrent data structure to a representation optimized for read-heavy
  workloads (\Cref{sec:eval}).
\end{itemize}

\section{Background and Related Work}\label{sec:background}

At first glance,
the problem we pose would seem to bear similarity to previous
work on {\em multi-word transactions} (MCAS or STM).
%
%
For instance, \citet{Harris:2002:PMC:645959.676137} propose
a way to perform multi-word CAS (MCAS).  However this approach is effective only
for small, fixed numbers of locations rather than large, lock-free data
structures---it must create a descriptor proportional to all addresses being
changed, which we cannot do for a data structure which grows during
transitioning, and would be inefficient in any case.
Software transactional memory~\cite{Herlihy:2003:STM:872035.872048} can be used
on dynamic-sized data to implement obstruction-free data structures
\cite{Herlihy:2003:OSD:850929.851942}.  However, STM's optimistic approach in
general cannot scale to large data structures.


\mypara{Snapshots}

Further, \citet{Afek:1993:ASS:153724.153741} give an algorithm
to archive atomic snapshots of shared memory, however it again works for fixed
locations and does not apply to dynamic data structures.  Furthermore, snapshot
alone cannot implement {\em transition} between lock-free data structures,
building a new structure based on the snapshot would lose the effects of ongoing
method calls.
However, the converse works: the transition algorithm we will present can be
used to obtain a snapshot for any lock-free data structures.


\mypara{Functional data structures}
Purely functional languages such as Haskell have a leg up when it comes to
snapshotting and migrating data.  This is because immutable data forms the {\em
  majority of the heap} and mutability status is visible both in the type system
and to the runtime.  Mutable locations are accessible from multiple threads few and
far between.
Indeed, this was recently studied empirically across 17 million lines of Haskell code,
leading to the conclusion that potentially-racing IO operations are quite rare
in existing Haskell code \cite{sc-haskell-draft}.
%
At one extreme, a single mutable reference is used to store a large purely
functional structure---this scales poorly under concurrent access but offers
$O(1)$ snapshot operations.

%
%
%
Some functional languages already include {\em freezable} data
structures~\cite{Leino:2008:FIF:1434628.1434650,Gordon:2012:URI:2384616.2384619}
that are initially mutable and then {become} immutable.  These include Haskell
vectors, Clojure transients, and IVars in Haskell or Id~\cite{monad-par,id}.
%
%
{But these approaches all assume that only a {\em single thread} writes the data
  while it is in its mutable state.}





\mypara{Adaptive data, prior work}

Previous work by Newton et. al.~\cite{adaptive-data-icfp15} leverages the
snapshot benefit of immutable data in a mutable container, presenting an
algorithm to transition from purely functional map implementations to fully
concurrent ones upon detecting contention (``heating up'').
%
%
The combined, hybrid data structure was proven to preserve
lock-freedom.
There are several drawbacks of this approach, however\footnote{
(1) the hybrid could only make one transition.  There was no way to
    transition back or transition forward to another implementation.
(2) it was restricted to map-like data with {\em set semantics}.
(3) it required storing {\em tombstone} values in the target
    (post-transition) data structure, which typically adds an extra level of
    indirection {\em to every element}.
%
 Our work removes these restrictions.
}.  Most importantly, the algorithm assumed a starting state of a pure data
structure inside a single mutable reference.
What about adapting from a lock-free concurrent data structure as the {\em
  starting state}?

\mypara{Example: Cooling down data}


{As a motivating example, consider a cloud document stored on a server that
  experiences concurrent writes while it is being created, and then
  read-only accesses for the rest of its lifetime.}
\new{Likewise, applications that use time series data (e.g., analytics)
  handle concurrent write-heavy operations, followed by a cool-down phase---a
  read-only workload, such as running machine learning algorithms.}
%
\nnew{But in order to support these scenarios, we first need to introduce our
  basic building block---freezable references.}

\rnfloat{TODO: We need to clarify our restrictions up front, and walk people through how
  to convert lock free objects to our last-operation-commits representation.}

\rnfloat{TODO: Also, address generality and shrink the scope of the claims a bit.}

\section{Prerequisite: Freezable IORefs}
\label{sec:iorefandhybrid}

In this section, we describe \new{the interface to a freezable reference. This
  API could be implemented in any language, but because we use Haskell for our
  experiments (\Cref{sec:eval}), we follow the conventions of the Haskell data
  type \il{IORef}.}



\begin{snippet}
  newIORef   :: a            -> IO (IORef a)
  readIORef  :: IORef a      -> IO a
  writeIORef :: IORef a -> a -> IO ()
\end{snippet}

\new{Ultimately, we need only one new bit of information per IORef---a frozen
  bit.  The reference is frozen with a call to \il{freezeIORef}:}
\begin{snippet}
  freezeIORef :: IORef a -> IO ()
\end{snippet}

After this call, any further attempts to \il{writeIORef} will raise an
exception. In our formal treatment, we model this \new{by treating exceptions as
  values, using a sum type with the} @Left@ \new{value indicating an
  exception. This is a straightforward translation of the} @EitherT@ or
@ExceptionT@ \new{monad transformers in Haskell. However, in our implementation,
  {we use Haskell's existing exception
    facility}~\cite{haskell-imprecise-exceptions}. Internally, freeze must be
  implemented with compare-and-swap (CAS) which requires a retry loop, a topic
  discussed in \Cref{sec:atomic-freezing}.}


\mypara{CAS in a purely functional language}
There are some additional wrinkles in exposing CAS in a purely functional
language, which, after all, normally does not have a notion of pointer equality.
The approach to this in Haskell is described in \citet{adaptive-data-icfp15},
and reviewed in ~\cref{sec:tickets}.

\section{A lifted type for Hybrid Data}\label{sec:lifted}

\new{Next we build on freezable references, by assuming a starting lock-free
  data structure, A, which uses {\em only} freezable and not raw references internally.
  Because of the matching interface between freezable and raw references, this
  can be achieved by changing the module import line, or any programming
  mechanism for switching the implementation of mutable references within A's
  implementation.}

We next introduce a datatype that will internally convert between the A and B structures, with types @DS_A@ and @DS_B@ respectively.
\new{We assume each of these data constructors is of kind \il{* -> *}, and that
  \il{(DS_A t)} contains elements of type \il{t}.}
The lifted type for the hybrid data structure is given in
\Cref{fig:l_type}. There are three states: @A@, @AB@ and @B@. States @A@ and @B@
indicate that the current implementation is @DS_A@ and @DS_B@ respectively, and
state @AB@ indicates that transition has been initiated but not finished yet.

\begin{figure}
  \begin{snippet}
   import Data.IORef -- Original, non-freezable refs.

   data State a = A  (DS_A a)
                | AB (DS_A a) (DS_B a)
                | B  (DS_B a)
   type DS a = IORef (State a)
  \end{snippet}
  \caption{Lifted type for a hybrid A/B structure}
  \label{fig:l_type}
\end{figure}

We also require two additional functions on @DS_A@:
\begin{lstlisting}[style=snippet]
  freeze  :: DS_A a -> IO ()
  convert :: DS_A a -> DS_B a -> IO ()
\end{lstlisting}


@freeze@ must run @freezeIORef@ on {\em all} @IORef@s inside @DS_A@, and then
@convert@ constructs a @DS_B@ from a frozen @DS_A@.
\new{While freeze is running, any write operations to the data structure may or may
not encounter a frozen reference and abort.}

Using the @freeze@ and @convert@ functions, we can construct a {\em transition}
function, as shown in \Cref{fig:tran}. @transition@ first tries to change the
state @A@ to @AB@, then it uses @freeze@ to convert all @IORef@s in @DS_A@ to
the frozen state, followed by a call to @convert@ to create @DS_B@ from
@DS_A@. Finally, @transition@ changes the state to @B@, indicating completion.
\nnew{Later, we will discuss {\em helping} during freeze and convert.
  Surprisingly, while helping algorithms usually require all participants to
  respect a common order, the freeze phase allows all participants to freeze A's
  internals in arbitrary order.}

\new{Because \il{convert} takes B as an {\em output} parameter, the
  implementation retains flexibility as to how threads interact with this global
  reference.  In this section, only a single thread populates B, but in the next
  section, multiple calls to convert will share the same output destination, at
  which point threads can race to install their privately allocated clone, or
  cooperate to build the output.  The specific strategy for \il{convert} is
  data structure specific.  And \il{convert}, like \il{freeze}, is an {\em
    prerequisite} to the hybrid algorithm.}

\begin{figure}
  \begin{snippet}
  transition :: DS a -> IO ()
  transition m = do
    tik <- readForCAS m
    case peekTicket tik of
      A a -> do
        b <- DS_B.new
        (flag, tik') <- casIORef m tik (AB a b)
        if flag then do
          DS_A.freeze a
          convert a b
          casLoop tik' b
        else return ()
      AB _ _ -> return ()
      B _ -> return ()
    where
      casLoop tik b = do
        (flag, tik') <- casIORef m tik (B b)
        unless flag $ casLoop tik' b
  \end{snippet}
  \caption{General template for a transition function, waiting/blocking version}
  \label{fig:tran}
\end{figure}
\par

\subsection{First hybrid algorithm: without Lock-freedom}


In this section we describe how to build a hybrid data structure that does not
preserve lock-freedom but has good performance assuming a fair scheduler.
We later describe a version which preserves lock-freedom in~\cref{sec:helping}
via small modifications to the algorithm in this section.

We assume that @DS_A@ and @DS_B@ only have two kinds of operations, @ro_op@ and
@rw_op@, where @ro_op@ indicates a read-only operation that doesn't modify any
mutable data, and @rw_op@ indicates a read-write operation that may modify
mutable data. We give the implementations for the hybrid versions of @ro_op@ and
@rw_op@ in \Cref{fig:ro_op,fig:rw_op}.

\begin{figure}
  \begin{snippet}
   ro_op :: a -> DS a -> IO b
   ro_op v r = do
     state <- readIORef r
     case state of
       A  a   -> DS_A.ro_op v a
       AB a _ -> DS_A.ro_op v a
       B  b   -> DS_B.ro_op v b
  \end{snippet}
  \caption{General template for a hybrid read-only operation.  The data
    structure is always available for read-only operations, even during
    transition.
  }
  \label{fig:ro_op}
\end{figure}

The @ro_op@ in \Cref{fig:ro_op} {\em does not mutate} any @IORef@, so it
needn't handle exceptions.  Rather, it merely calls the corresponding
function from @DS_A@ or @DS_B@.  It also has the {\bf availability property that
  it can always complete quickly, even while transition is happening}. This is
somewhat surprising considering that transitioning involves freezing references
in an arbitrary order, and partially completed (then aborted) operations on the
@DS_A@ structure.  But as we will see, the strong lock-freedom property on
@DS_A@ makes this possible.

\begin{figure}
  \begin{snippet}
 rw_op :: a -> DS a -> IO ()
 rw_op v r = do
   state <- readIORef r
   case state of
     A  a   -> handler $ DS_A.rw_op v a
     AB _ _ -> do yield; rw_op v r
     B  b   -> DS_B.rw_op v b
   where
     handler f =
       runExceptT f >>= \c -> case c of
         Left e  -> do yield; rw_op v r
         Right a -> return a
  \end{snippet}
  \caption{General template for a hybrid read-write operation.  Waiting/blocking version.
  }
  \label{fig:rw_op}
\end{figure}

For @rw_op@ in \Cref{fig:rw_op}, if the current state is one of @A@ or @B@, it
calls the corresponding function from @DS_A@ or @DS_B@. If the state is @AB@,
{we spin until the transition is completed}, which is the key action that
loses lock-freedom by waiting on another thread.
Since @rw_op@ changes the @IORef@, we also need to handle a
\new{CAS-frozen exception, which is the job of \il{handler} and \il{runExceptT}.}
This exception can only be raised \new{after transition begins}, so the method retries
until the transition is finished.

\new{Exception handling is key to the hybrid data structure.  But the most
  surprising feature of this algorithm is not directly visible in the code: {\bf
    the original algorithm for data structure A need not have {\em any awareness of
    freezing}}.  Rather, it is an arbitrary lock-free data structure with its
  mutable references hijacked and replaced with the freezable reference type.
  The algorithm for \il{rw_op} works in spite of CAS operations {\em aborting} at
  arbitrary points.}

This is a testament to the strength of the lock-freedom guarantee on A.
\nnew{In fact, this property of lock-free algorithms---that intermediate states
  are valid after a halt or crash---is known in the folklore and was remarked
  upon informally by Nawab et. al.~\cite{lock-free-crash-resilience} in the
  context of crash recovery in non-volatile memory.  We believe this paper is
  the first that formalizes and proves this proposition.}


{Finally,} note that the lifted type @DS@ can easily be extended to more than three states,
for example, we can transition back from @DS_B@ to @DS_A@, or transition to a
different data structure @DS_C@.  Thus it is without loss of generality that we
focus on a single transition in this paper, and on representative abstract
methods such as @rw_op@.  \new{Our evaluation, in \Cref{sec:eval}, uses
  a concrete instantiation of the two-state A/B data structure to demonstrate
  the method.}


\mypara{Progress guarantee}
Even though this version of the algorithm requires that threads wait on the
thread performing the freeze, it retains a starvation-freedom guarantee.  The
freezing/converting thread that holds the ``lock'' on the structure
monotonically makes progress until it releases the lock, at which point waiting
@rw_op@s can complete.



\section{Helping for Lock-freedom} \label{sec:helping}




%

\new{The basic algorithm from the previous section is an easy starting point. It
  provides the availability guarantee and arbitrary-freezing-order properties by
  virtue of A's lock freedom and in spite of the hybrid's lack of the same.
  Now we take the next step, using the established concept of {\em
    helping}~\cite{Agrawal:2010:HLF:1693453.1693487} to modify the hybrid
  algorithm and achieve lock-freedom.}
Specifically, threads help each other to accomplish the freeze and convert
steps.  \Cref{fig:rw_op_lf} gives the new algorithm for wrapping each method of
a data structure to enable freezing and transitioning to a new representation.
Again, the choice of when to transition is external, and begins when a client
calls @transition@.  \new{The modified, lock-free, transition function is shown
  in \Cref{fig:tran_lf}.}

\new{In both these figures, the unaltered lines are shown in gray. Also, they
  include a \il{method} wrapper form. This has no effect during execution, but
  is present for administration purposes as will be explained in
  \Cref{sec:correctness}.}

\new{Only two lines of code in the body of \il{rw_op} have changed.  Now any
  method that attempts to run on an already-transitioning data structure invokes
  transition itself, rather than simply retrying until it succeeds.  Likewise
  for methods that encounter a CAS-on-frozen exception.}
Intuitively, once transitioning begins, it must complete
in bounded time because all client threads that arrive to begin a new operation instead {\em help}
with the transition.  Thus even an adversarial schedule cannot stall
transitioning.
%
We target this refinement of the algorithm in our proofs of correctness in
\cref{sec:correctness}.

\new{Finally, note that the lock-freedom progress guarantee holds whether
  multiple threads truly cooperate in freezing, or simply start $N$ redundant
  freeze operations. This is an implementation detail within the \il{freeze}
  function.  Clearly, cooperative freezing can improve constant factors compared
  to redundant freezing.}
%
The same argument applies for the @convert@ function, and in
\Cref{sec:cooperative} we give an algorithm for such cooperation.


\section{Correctness} \label{sec:correctness}


We propose an abstract machine that models arbitrary lock-free data structures'
methods interacting within a shared heap. In this section, we use these abstract
machines in conjunction with operational style arguments\footnote{Standard in the
  concurrent data structure
  literature~\cite{hendler-stack-2004,split-ordered-list,herlihy1990linearizability}} to show that correct, lock-free starting and
ending structures ({\em A} and {\em B}) yield a correct, lock-free hybrid data
structure \new{using the algorithm of the previous section}.
\nnew{One of the surprising aspects of this development is that our proof goes
  through {\bf without any assumptions about a canonical ordering} in
  regarding ``helping'' for freeze and convert---indeed, our
  implementation in~\cref{sec:cooperative} uses randomization.}


\begin{figure}
  \begin{code}
@rw_op :: a -> DS a -> IO ()
rw_op v r = @method@ { do
  state <- readIORef r  (*@\label{l_rw_read}@*)
  case state of
    A  a -> handler $ DS_A.rw_op v a@  (*@\label{l_rw_commit_A}@*)
    AB _ _ -> do transition r (*@\label{l_rw_trans}@*)
                 @rw_op v r  (*@\label{l_freeze3}@*)
    B  b -> DS_B.rw_op v b@  (*@\label{l_rw_commit_B}@*)
  @}
  where
    handler f =
      runExceptT f >>= \c -> case c of
        Left e  -> do @transition r@ (*@\label{l_rw_trans2}@*)
                      rw_op v r    (*@\label{l_rw_trans3}@*)
        Right a -> return a@
  \end{code}
  \caption{Lock-free, hybrid read-write operation}
  \label{fig:rw_op_lf}
\end{figure}

\begin{figure}
  \begin{code}[mathescape=true]
@transition :: DS a -> IO ()
transition m = @method@ { do
  tik <- readForCAS m
  case peekTicket tik of
   A a -> do
     b <- DS_B.new
     (flg,tik') <- casIORef m tik (AB a b) (*@\label{l_tran_cas}@*)
     if flg
       then do
         DS_A.freeze a@ (*@\label{l_freeze1}@*)
         convert a b (*@\label{l_convert1}@*)
         casIORef m tik' (B b) (*@\label{l_tran_casB1}@*)
       else transition m
   AB a b -> do
     DS_A.freeze a (*@\label{l_freeze2}@*)
     convert a b (*@\label{l_convert2}@*)
     casIORef m tik (B b)@ (*@\label{l_tran_casB2}@*)
   B _ -> @noop m@ (*@\label{l_tranB}@*)
  }@
  \end{code}
  \caption{Lock-free transition function}
  \label{fig:tran_lf}
\end{figure}

\subsection{Formal Term Language}

\newcommand{\bnfdef}{\ensuremath{::=\;}}
\newcommand{\alt}{\ensuremath{\;|\;}}

\newcommand{\Right}[1]{\ensuremath{\textsf{Right} \; #1}}
\newcommand{\Left}[1]{\ensuremath{\textsf{Left} \; #1}}

\newcommand{\fst}[1]{\ensuremath{\textsf{fst} \; #1}}
\newcommand{\snd}[1]{\ensuremath{\textsf{snd} \; #1}}

\newcommand{\method}[2]{\ensuremath{\textsf{method}_{#1}\,\{\; #2 \; \}}}

\newcommand{\case}[3]{\ensuremath{\textsf{case} \; #1 \;
                      \textsf{of} \,
                      \Left x \rightarrow #2; \,
                      \Right x \rightarrow #3}}

\newcommand{\bind}[2]{\ensuremath{#1 \; \texttt{>>=} \; #2}}
\newcommand{\ret}{\ensuremath{\textsf{return}\;}}

\newcommand{\Int}{\ensuremath{\textsf{Int}}}
\newcommand{\cas}{\ensuremath{\textsf{casRef}}}
\newcommand{\frz}{\ensuremath{\textsf{frzRef}\;}}
\newcommand{\newrefName}{\ensuremath{\textsf{newRef}}}
\newcommand{\newref}{\ensuremath{\newrefName\;}}
\newcommand{\Ref}{\ensuremath{\textsf{Ref}\;}}

\newcommand{\getroot}[1]{\ensuremath{root(#1)}}

\newcommand{\Locs}{\textsf{Locations}}
\newcommand{\loc}{\ensuremath{\ell}}

\newcommand{\Vals}{\textsf{Values}}

\newcommand{\fix}{\ensuremath{\textsf{fix}\;}}

\newcommand{\IO}{\ensuremath{\textsf{IO}\;}}

\newcommand{\unit}{\ensuremath{\texttt{()}}}

\newcommand{\Type}{\ensuremath{T}}

\newcommand{\conv}[1][{}]{\rightarrowtriangle^{#1}}

\newcommand{\subst}[3]{#1[#2/#3]}

In our abstract machine we use a small, call-by-need term language
corresponding to a subset of Haskell.  Its grammar is given in
\Cref{fig:grammar}.

\mypara{Desugaring}

In our code figures, such as \Cref{fig:rw_op_lf}, we took the liberty of
including features and syntactic sugars that have a well-understood translation
into the core language of \Cref{fig:grammar}.  Namely, we allow:

\begin{compactitem}
\item \il{do} notation for monadic programming.
\item The {\em Exception} monad, implemented with binary sums.
\item N-way sum types, which are mapped onto binary sums.
\item Top-level recursive bindings, using a \fix combinator.
\item Reads as always-fail CAS operations, described in \Cref{sec:cas}.
\item Read and CAS operations on references containing compound data using {\em
    double-indirection} (\Cref{sec:double-indirection}).
\end{compactitem}

\noindent
There are also two distinctions in our language that are {\em not} present in
standard Haskell, but express invariants important to our proofs.
First, the ``\method{}{}'' syntax is used to assert that a given monadic action is
linearizable
and that the {\bf {\em last} memory operation in that action is its
  linearization point}.  To illustrate how this works, it is useful to compare
against the atomic section used in some languages.
An atomic section implies runtime support for mutual exclusion, whereas the
\textsf{method} form only captures a semantic {\em guarantee}.

\label{sec:last-instruction}
\nnew{The last-instruction convention may seem restrictive.  For example, in the
  lazy lists algorithm of Harris~\cite{lazy-lists-Harris:2001}, methods
  linearize and then subsequently perform cleanup, likewise for skip-lists which
  opportunistically update the indexing structure after they insert. The key
  insight is that {\bf post-linearization actions are necessarily
    non-semantic}. They are cleanup actions which become irrelevant after
  transition occurs. We model a method $m$ with a linearization point in the
  middle, as two methods, $m_1$, $m_2$, where the first performs the semantic
  change, and the second the cleanup. After transition, $m_2$ may logically be
  rescheduled, but it is just a NOOP in the new representation.}


As a consequence of requiring that methods finish on a memory operation, we
include a dummy memory operation on line \ref{l_tranB}, as a placeholder for the
@transition@ method to linearize.  That is, \il{noop = readForCAS}, whereas in
practice it could just as well be \il{noop _ = return ()}.

The second distinction present in our formal language (but not Haskell) is the
{\em Actions} nonterminal.  {Actions} include all the same forms as expressions,
plus operations on references.  Separating actions and expressions ensures that
client programs only call memory operations from {\em within} a \method{}{}
form.

Finally, note that for a compound method built from other methods --- such as
our hybrid data operations --- the norm is to call inner methods in {\em tail
  call} position.  Thus the linearization point of the inner method becomes the
linearization point of the outer method \new{(as in \Cref{fig:semantics1})}.

Any methods called in non-tail position are unusual because they must complete
before the linearization of the containing method.  That is, they must have no
{\em semantic effect}.  Indeed, this is precisely the case with the
\il{transition} method, which is called in non-tail position in several places,
including \Cref{l_rw_trans}.

\newcommand{\headingtweak}{}
{\small
  \begin{figure}
\[
\begin{array}{ll}
  x \in \textsf{Variables}   & \qquad n \in \textsf{Integers} \\
  m \in \textsf{MethodNames} & \qquad \loc \in \Locs \\
\end{array}
\]
 \begin{gather*}
\begin{alignat*}{2}
  &\textsf{Expressions} \\
  &\qquad e \bnfdef && x \alt v \alt \lambda x.e \alt e\;e \alt \fix e \alt
                       (e,e) \alt \fst{e} \alt \snd{e} \alt \\
                  & && \case{e}{e}{e} \alt
                       \Right{e} \alt \Left{e} \alt \\
                  & && \bind{e}{e} \alt \ret e \alt
                       \method{m}{a} \\
  & \textsf{\headingtweak{}Actions} \\
  &\qquad a \bnfdef && x \alt v \alt \lambda x.a \alt \dots \alt \\
                  & && \newref a \alt \cas(a,a,a) \alt \frz a \\
  &\Vals \\
  &\qquad v \bnfdef && n \alt \loc \alt \lambda x.e
                       \alt (e,e) \alt \Right{e} \alt \Left{e}
                      \\
  &\textsf{\headingtweak{}Types} \\
  &\qquad \Type \bnfdef && \Int
            \alt \Type \rightarrow \Type
            \alt (\Type,\Type)
            \alt \Type + \Type \alt
            \Ref \Type
            \alt \unit
            \alt \IO \Type
            \\
  &\textsf{\headingtweak{}Contexts} \\
  &\qquad E \bnfdef && [\;]
                     \alt E\;e \alt
                            \new{\method{m}{E} \alt}  \\
                  & && \new{\case{E}{e}{e}}
                   \alt
                       \bind{E}{e}
\end{alignat*}
\end{gather*}
\caption{{Core language grammar}}

\label{fig:grammar}
\end{figure}}


\subsection{Abstract Machine}

A data structure comes with a set of
methods $M = \{ (m_{1},e_{1}) , \allowbreak (m_{2},e_{2}) , \allowbreak \dots (m_{n},e_{n}) \}$,
\new{which maps method names, $m_i$, onto method implementations,
  which are expressions of the form: \\
  $\lambda {x} . \allowbreak\method{m_i}{e_{bod}}$.}
\new{Note here that methods take a {single} argument---the data structure
  to act on.  In our formal model, we don't include other arguments, because
  each method can be specialized to include these implicitly.
  For example, a concurrent set of finite integers can be modeled with
  with methods: \\
  \sloppy
  $\{ \tt{insert\_1}, \tt{insert\_2}, \dots \allowbreak
    \tt{remove\_1}, \tt{remove\_2}, \dots \}$.
Where it is not ambiguous, we use $m_i$ to interchangeably refer to the named
method as well as the expression bound to it in $M$.}

%
%

\fussy
We use an abstract machine to model a data structure interacting with a set of
{\em client threads}. Thread IDs are drawn from a finite set $\threadids{} = \{
\tau_1, \tau_2 \dots , \tau_k \}$, and expressions $\{ e_1 , e_2 , \allowbreak
\dots e_k \}$ evaluated by those clients.
\new{Clients can perform {\em arbitrary} local computation, but modify the
  shared memory only via method invocations.  Moreover, clients are
  syntactically restricted to include only method expressions drawn from $M$.  This
  facilitates defining {\em traces} of client method calls as merely lists of
  method names.}

%


%

\mypara{Pure Evaluation Semantics}

The semantics given in \Cref{fig:semantics1} are standard, with the exception of
the rule for eliminating nested \method{}{} syntax.  Eliminating nested methods
is necessary to reduce expressions to a form where the judgments in
\Cref{fig:semantics2} can fire.  The rule we choose preserves the label of the
outer method, discarding the inner one.  Note that the combination of this rule
and \textsc{MethodFinish} implies that only {\em top level} method calls within
clients are traced.  First, modeling inner methods would significantly
complicate matters.
Second, the top-level methods, which are the ones named in $M$, are the subject
of the lock-freedom guarantees \new{we will discuss shortly}.

\mypara{Shared Memory Operations}

The shared memory operations performed by methods are {\em compare-and-swap} and
{\em freeze} operations, \new{resulting from the \cas{} and \frz{} forms, respectively:}
\[
\begin{array}{ccc}
  MemOps = Cas \cup Frz && \\
  Frz = \lbrace frz(\loc) ~|~ \loc \in \Locs \rbrace && \\
  Cas = \lbrace {cas(\loc,v_{old},v_{new})} ~|~ \loc \in \Locs, v \in \Vals \rbrace
\end{array}
\]
\new{These operations mutate a central heap, $H$, which stores the contents of
  all references.  We further assume a designated location $\getroot{H} =
  \loc_{root}$, storing the data structure.  It is this location that is, by
  convention, the argument to methods.
  }

With these prerequisites, we define the configuration of an
abstract machine, $\sigma$, as follows:

\begin{definition}[Abstract machine]
  $\sigma^A = (t,H,F,M,\Trace,\Pool)$ is a configuration of data structure A which consists of:
  \begin{itemize}
  \item the current time, $t : \mathbb{N}$,
  \item a heap state containing A's representation, \\
    {$H : \Locs \rightarrow \Vals$
      },
  \item
    a finite set $F$ of frozen addresses, $F \subset \Locs$,

  \item the set of methods, $M$, and

  \item \new{a trace of completed methods, containing pairs of method name and
    return value: $\Trace : [ (\textsf{MethodNames} , \textsf{Values}) ]$,}

  \item an active method pool $\Pool : \threadids{} \rightarrow e$, which is a finite map
    from thread IDs to the current expression which is running on the client thread
  \end{itemize}
\end{definition}

\new{Here we assume discrete timesteps and focus on modifications to shared
  memory.  Pure, thread-local computation is instantaneous; it does not
  increment the time.  Further, the semantics of the language is split into pure
  reductions (\Cref{fig:semantics1}) and machine-level IO actions
  (\Cref{fig:semantics2}).  Equivalently, one can deinterleave these reduction
  relations and view the pure reduction as happening first and producing an
  infinite tree of IO actions~\cite{concurrent-haskell-popl96}.  For our
  purposes, the differences are immaterial and we choose the former option.}

%
%
%

\begin{definition}[Initial Conditions]
For each data structure A, we assume the existence of an initial heap state
$H_{init}^A$. Given a map from thread ID to client programs, $\Pool$, an initial configuration is
  $\sigma_{init}^A(\Pool) = (0,H_{init}^A, \ouremptyset, M^A, [], \Pool )$.
\end{definition}

An initial heap state represents, e.g., an empty collection data structure or a
counter initialized to zero.
The heap is a finite map, and
we refer to the size of the heap $|H|$.
%
%
\new{Note that both \cas{} and \frz{} operate only on {\em bound} locations
  ($\loc \in dom(H)$), whereas
\newrefName{} binds fresh locations in the heap,  as shown in \Cref{fig:semantics2}.}


In summary, $\sigma$ contains a data structure's implementation and an active
instantiation of that structure, plus a representation of executing clients. One
final missing piece is to specify the {\em schedule}, which models how an
operating system selects threads at runtime.

\begin{definition}[Schedule]
  A schedule $\pi$ is a function $\mathbb{N} \rightarrow \threadids{}$
  that specifies which thread to run at each discrete timestep.
\end{definition}

Together, $\sigma$ and $\pi$ provide the context needed to reduce a
configuration as we describe in the next section.


\subsection{Stepping the Machine}

An evaluation step
$\rightarrow_{\pi}$
is a relation that depends on the schedule $\pi$.
\Cref{fig:semantics2} gives the dynamic semantics.

\subsubsection{Semantics of CAS}\label{sec:cas}


Where $\cas(e,e,e)$ is a concrete term, $cas$ is the abstract operation that
results from its evaluation.  We write $H[cas]$ to indicate the heap $H$ updated
with a successful CAS operation; $H[\loc \mapsto v]$ for finite map update,
replacing any previous value at location $\loc$. Moreover, we use
$v_{res} \leftarrow H[cas]$ to bind the name $v_{res}$ to the value returned by
the CAS.


\new{The updated heap after a single CAS operation, $H' = \allowbreak{}H[cas(\loc,\allowbreak{}old,new)]$, is
given by:}
\[
\begin{array}{l}
H' =
\begin{cases}
  H                   & \text{if } \loc \in F \\
  H                   & \text{if } \loc \notin F \wedge H(\loc) \neq old \\
  H[\loc \mapsto new] & \text{if } \loc \notin F \wedge H(\loc) = old \\
\end{cases}
\end{array}
\]

\noindent
\new{Likewise, the return value $v_{res} \leftarrow H[cas(\loc,old,new)]$, is:}
\[
\begin{array}{l}
v_{res} =
\begin{cases}
  \Left  H(\loc) & \text{if } \loc \in F \\
  \Right H(\loc) & \text{if } \loc \notin F \wedge H(\loc) \neq old \\
  \Right new     & \text{if } \loc \notin F \wedge H(\loc) = old \\
\end{cases}
\end{array}
\]


We call the last case a {\em successful} CAS, as it
mutates the heap.  The first two cases are both failures.
The $\Left$value communicates
to the caller that the location is frozen and all future CAS attempts will fail.
This is a ``freeze exception'', and is propagated by the containing method as an
exceptional return value (e.g. \Cref{fig:rw_op_lf}).


\new{Finally, when desugaring Haskell code to our formal language, we omit
  \il{readForCAS} as a primitive action.  Rather, we define a read as a CAS with
  an unused dummy value, which will never succeed:}
\begin{snippet}[mathescape=true]
  readForCAS e == case (*@\cas(\text{e},d,d) @*) of
                   Left x -> x
                   Right x -> x
\end{snippet}
\new{This dummy value is type specific; for our uses in this paper, it can be
  simply an otherwise unused location, $\loc_0$.}

\paragraph{CAS on compound types}

Because our formal model only allows CAS on addresses containing a single
location or integer (a ``word''), we need to encode atomic swaps on pairs just
as we do in a real implementation---using an extra level of indirection and
CASing the pointer.  This double-indirection encoding is explained in more
detail in~\cref{sec:double-indirection}.


\subsubsection{Machine-level Operational Semantics}


\new{\Cref{fig:semantics2} gives the operational semantics.  Here we explain
  each judgment of the $\rightarrow_{\pi}$ relation which covers all primitive,
  atomic IO actions.}
%
%
\begin{itemize}


\item {\textsc{New}}: Create a new reference in the heap, initializing it with
  the given value.


\item {\textsc{Frz}}: An atomic memory operation that freezes exactly one address.

\item {\textsc{Cas}}: A thread issues a CAS operation that either succeeds, writing a
  new value to the heap, or fails, having no effect on the heap but returning
  the most recent value to the calling method (i.e. serving as a {\em read}
  operation).

\item {\textsc{MethodFinish}}: The final expression must be a CAS which is also
  the linearization point of that method.  A finish step updates the trace of
  the abstract machine by adding the method name and return value.

\item {\textsc{MethodPure}}: Pure computation only updates the expression in the pool.


\end{itemize}

\mypara{Atomic freezing} \label{sec:atomic-freezing}
{Note that $frz$ {always succeeds in this model, in one reduction step.  In
  practice, this is typically implemented as a retry loop around a
  machine-level}
  CAS instruction {(e.g.  \il{freezeIORef} in \Cref{fig:freezeIORef}).}
  We could just as well model $frz$ in the same way as
  CAS---taking an expected old value and succeeding or failing.  But it is not
  necessary.  In practice, any number of contending $frz$ and $cas$ operations
  still make progress, and we have no need of modeling physical machine states
  where $frz$ fails.
}
(Further, our algorithm only employs $frz$ instructions during the $AB$ state,
during which there are a bounded number of non-$frz$ memory operations.)







{\small
  \begin{figure}
  \begin{mathpar}


  \inferrule [CAS]
             {\Gamma \vdash e_1 : \Ref T \\
               T = Int \;\vee\; T = \Ref T_2 \\
               \Gamma \vdash e_2, e_3 : T}
   {\Gamma \vdash \cas(e_1,e_2,e_3) : \IO (T + T)} \and

  \end{mathpar}
  \caption{Static semantics, standard rules omitted.  Note that the return type
    of CAS is a sum type admitting failure.  Here, failure occurs when the
    address is {\em frozen}.}
  \label{fig:types}
\end{figure}}

{\small
\begin{figure}
  \begin{mathpar}

    \method{m_1}{\method{m_2}{ e }} \conv \method{m_1}{ e }

    \bind{\ret e_1}{e_2} \conv e_2 \; e_1

    (\lambda x.e_1) \; e_2 \conv e_2[e_1/x]

    {\fst (e_1,e_2)} \conv e_1

    {\snd (e_1,e_2)} \conv e_2



    (\case{\Left e_1}{e_2}{e_3})  \conv e_2[e_1/x]

    (\case{\Right e_1}{e_2}{e_3})  \conv e_3[e_1/x]

    {\fix (\lambda x.e)} \conv e[\fix (\lambda x.e)/x]

    \inferrule*{e \conv e'}
               {E[e] \conv E[e']}
  \end{mathpar}
  \caption{{
      Small step operational semantics for term language.
      This relation handles pure computation only; thus \cas{} and
      \frz{} do not reduce.}}
  \label{fig:semantics1}
\end{figure}}

{\small
  \begin{figure*}
  \begin{mathpar}


    \inferrule*
        [lab=New]
        {\pi(t) = \tau \\
          \Pool(\tau) = E[ {\method{m}{ \bind{\newref{v}}{e_1} }} ] \\
          \loc \notin dom(H)
        }
        {
          (t,H,F,M,\Trace,\Pool)
          \stepsto{\pi}
          (t+1,H[\loc \mapsto v],F,M,\Trace,\Pool[\tau \mapsto E[ {\method{m}{e_1 \; \loc}}]])
        }

    \inferrule*
        [lab=MethodFinish\ensuremath{(m)}]
        {\pi(t) = \tau \\
          \Pool(\tau) = E[{\method{m}{ \cas(\loc,v_1,v_2) }}] \\
          v_{res} \leftarrow H[cas(\loc,v_1,v_2)]
        }
        {
          (t,H,F,M,\Trace,\Pool)
          \stepsto{\pi}
          (t+1,H[\cas(\loc,v_1,v_2)],F,M,\Trace \doubleplus [(m , v_{res})], \Pool[\tau \mapsto E[v_{res}]])
        }

    \inferrule*
        [lab=Cas]
        {\pi(t) = \tau \\
          \Pool(\tau) = E[{\method{m}{ \bind{\cas(\loc,v_1,v_2)}{e_1} }}] \\
          v_{res} \leftarrow H[cas(\loc,v_1,v_2)]
        }
        {
          (t,H,F,M,\Trace,\Pool)
          \stepsto{\pi}
          (t+1,H[cas(\loc,v_1,v_2)],F,M,\Trace,\Pool[\tau \mapsto E[{\method{m}{e_1 \; v_{res}}}]])
        }

    \inferrule*
        [lab=Frz]
        {\pi(t) = \tau \\
          \Pool(\tau) = E[{\method{m}{ \bind{\frz \loc}{e_1} }}]
        }
        {(t,H,F,M,\Trace,\Pool)
          \stepsto{\pi}
          (t+1,H,F \cup \{\loc\},M,\Trace,\Pool[\tau \mapsto E[{\method{m}{e_1 \; \texttt{()}}}]])
        }





   \inferrule*
        [lab=MethodPure]
        {\pi(t) = \tau \\
          \Pool({\tau}) = E[e_1]\\
          e_1 \conv e_2}
        {
          (t,H,F,M,\Trace,\Pool)
          \stepsto{\pi}
          ({t},H,F,M,\Trace,\Pool[\tau \mapsto E[e_2] ])
        }

  \end{mathpar}
  \caption{{Dynamic semantics of a data structure abstract machine.}}
  \label{fig:semantics2}
\end{figure*}}

\subsubsection{Observable Equivalence}

Let $\twoheadrightarrow_{\pi}$ be the reflexive-transitive closure of $\rightarrow_{\pi}$. Further, let
${\sigma \twoheadrightarrow_{\pi}^n \sigma'}$ signify a reduction of $n$ steps.

\begin{definition}[Sequential execution]
  We use $\sigma[m_i]$ to denote the state
  that results from executing method $m_i$ {\em sequentially}, i.e.,
  on a thread $\tau$ under a schedule that selects only
  $\tau$ for future timesteps.
\end{definition}



\noindent
\new{Note that a state $\sigma$ may already include a set of executing clients.
  The syntax $\sigma[m_i]$ ignores these clients, introduces a fresh client
  thread id from \threadids, and proceeds to evaluate $m_i \; \loc_{root}$
  until it completes with \textsc{MethodFinish}.}

We abbreviate $\sigma[m_1][m_2] \dots [m_l]$ as $\sigma[m_1, m_2 \dots m_l]$. We refer to the results of such an
execution, $\results{\sigma[m_1, m_2,\allowbreak \dots m_l]}$, to mean the
sequence of \new{values $v_1 \dots v_l$} resulting
from method calls $m_1$ through $m_l$. Two configurations $\sigma_1$ and $\sigma_2$ are observably equivalent if any
sequence of method calls results in the same values.



\begin{definition}[Observable Equivalence]

  For abstract machines
  $\sigma_1 = (\_,\_,\_,M,\_,\_)$ and
  $\sigma_2 = (\_,\_,\_,M,\_,\_)$,
  ${\sigma_1 \approx \sigma_2}$ iff
  for all $\overline{m}$,
  $\results{\sigma_1[\overline{m}]} = \results{\sigma_2[\overline{m}]}$.
\end{definition}

So if $\sigma_1 \approx \sigma_2$, any internal differences in heap contents and
method implementations are not distinguishable by clients. Note, however, that
this definition considers only sequential method applications.  Thus we now
address {\em linearizability} which enables reasoning about concurrent
executions in terms of such sequential executions.
Linearizable methods are those that always appear as if they executed in
some sequential order, even if they didn't.  Commonly, in the definition of
lock-free algorithms, we identify {\em linearization points} as the atomic
instructions which determine {\em where} in the linear history the concurrent
method logically occurred.


As mentioned earlier, we assume that if a method has a {linearization point}, it
is the {\em last CAS instruction} in its trace.  This is without loss of
generality, as discussed in~\cref{sec:last-instruction}. Further, we
say that $\methods{\sigma_1 \twoheadrightarrow_{\pi} \sigma_2}$, is the sequence
$\overline{(m,v)}$
with one entry per \textsc{MethodFinish} reduction in the
reduction sequence.  Thus $\overline{(m,v)}$ corresponds to a subset of the trace,
$\Trace_2 - \Trace_1$.
This counts only methods that completed (linearized) after $\sigma_1$, up to and
including $\sigma_2$.


\begin{definition}[Linearizability]
  A data structure is linearizable iff, for all initial states $\sigma_{init}$ (including all client workloads),
  $\sigma_{init} \twoheadrightarrow \sigma'$ implies that the sequential application of methods
  $\overline{m} = \methods{\sigma_{init} \twoheadrightarrow \sigma'}$, yields the same observable result
  as the concurrent execution.
  That is:
    $\sigma_{init}[\overline{m}] \approx \sigma'$.
\end{definition}


\begin{definition}[Lock-Freedom]
  A data structure @A@ is lock-free iff there exists an upper bound function $f_A$, defined
  as a function of heap size $|H_A|$,
  such that, for all configurations $\sigma^A = (\_,H_A,\_,\_,\_,\_)$ and all schedules $\pi$, any reduction of $f_A(|H_A|)$
  steps completes at least one method.

  That is:
  $|\methods{\sigma_A \twoheadrightarrow_{\pi}^{f_A(|H_A|)} \sigma_A'}| \geq 1$.

\end{definition}

\subsubsection{Hybrid data structure}
\label{sec:hybrid}

The hybrid data structure described in \cref{sec:lifted} and
\cref{sec:helping} is built from two lock-free data structures, A and B.
%
A and B must share a common
set of (linearizable) methods $M$, and we assume that the data
structures are behaviorally indistinguishable. That is, for any
$\sigma^A_{init}(\Pool)=(0,H_{init_A},\ouremptyset,M,[\:],\Pool)$ and
$\sigma^B_{init}(\Pool)=(0,H_{init_B},\ouremptyset,M,[\:],\Pool)$,
we assume that ${\sigma^A_{init} \approx \sigma^B_{init}}$.

\sloppy
The hybrid data structure is modeled by the abstract machine
$\sigma^{Hybrid} = (t,H,F,M^+,\Trace,\Pool)$ where
$M^+ = M^{Hybrid} \cup \{transition\}$. Here, the set $M^{Hybrid}$ has a
one-to-one correspondence with the public interfaces $M^A$ and $M^B$ of data
structures A and B respectively, \new{and it shares the same method labels}.
But we extend the set of methods with an externally visible one that initiates
the $A \rightarrow B$ transition. @transition@ (\Cref{fig:tran_lf}) is a
full-fledged method in the sense that it has a linearization point and counts
towards progress vis-a-vis lock-freedom.



\fussy
We assume that locations, \loc, used in heaps $H_A$ are disjoint from addresses
used by $H_B$.
%
Thus A and B data structures can be copied directly into the hybrid heap $H$,
and coexist there.
Also, the root, $\getroot{H} = \loc_{hy}$ must ensure $\loc_{hy} \notin dom(H_A)
\cup dom(H_B)$.  Further, the hybrid structure points to a small amount of
extra data encoding the lifted data type shown in
\Cref{fig:l_type}.  Specifically, $\loc_{hy}$ contains a double-indirection
pointing to a sum type $A + (A + B)$, which is the binary-sum encoding of the
three-way A/AB/B datatype of \Cref{fig:l_type}.


This memory layout grants the
ability to {atomically} change {\em both} the state (A/AB/B) and the data
pointer by issuing a CAS on $\loc_{hy}$.
This capability is used by the code in \Cref{fig:l_type,fig:tran_lf,fig:rw_op_lf}. These
figures also show how to define each method, $m_i^{Hybrid}$
in terms of the originals, $m_i^{A}$ and $m_i^{B}$,
using the lock-free hybrid algorithm.
Further, we also use the functions, @freeze@ and @convert@, provided as by the
data structure A.
%
Below we describe the formal requirements on the terms that implement these two functions.





\mypara{Freeze function}


Given $\sigma^A = (t,H_A,F,M,\Trace,\Pool)$, @freeze@ is a function (not a
method) that will freeze all used locations in $H_A$ with $frz$ memory
operations. In fact, freeze is the {\em only} code which issues $frz$ operations
in our design. When the freeze function completes, it holds that $\forall \loc
\in dom(H_A) [\loc \in F ]$. We assume the existence of a freeze function for
the starting structure $A$ (but not $B$).
We further assume an upper bound on the time steps required for freeze to complete, $f_{frz}(|H_A|)$.

In our algorithm, @freeze@ is only used while the hybrid is in the @AB@
state. Thus if $\sigma \rightarrow \sigma'$ is a reduction that performs the
last operation in @freeze@,
then at all
later  points within the @AB@ state---that is all states reachable from
$\sigma'$ in which $\loc_{hy}$ is bound to an AB value---a
\cas{} {\em must} return a frozen exceptional value, $\Left v$.


There is {\bf no requirement} on the order in which @freeze@ freezes heap locations.
This may be surprising, because of the possibility that client threads could
allocate fresh locations faster than the freezing function can ice them.
However, our algorithm has a global property bounding the number of
\il{DS_A.rw_op} method calls that can execute after the transition to state
@AB@---\new{bounding it simply to the number of threads.  This in turn provides
  a} bound on the steps spent executing such methods.  That is, each
\il{DS_A.rw_op} takes a bounded number of steps because of the lock-freedom
assumption on data structure $A$.
Thus the termination of @freeze@ follows from the bounded number of fresh heap
locations that can be added to A's address space during the @AB@ state.
%

\mypara{Conversion function}

We assume a conversion function for converting $H_A$ heap representations to
$H_B$.  @convert@ is callable by methods and is implemented by reading locations
in A's portion of the address space and creating a fresh structure in B's
address space.

The requirement on @convert@ is that if it were lifted into its own method and applied to a state of A, then it
yields an observably equivalent state of B. That is,
$\sigma^A_{init}[\overline{m_A}]\allowbreak[convert] \approx \sigma^B_{init}[\overline{m_B}]$, where $m_A \equiv m_B$
are equivalent methods in the respective machines.
As with @freeze@, we assume an upper bound as a function of the heap size,
$f_{conv}(|H_A|)$, for the time steps necessary to complete conversion.

\subsection{Lock-freedom}

The full proof of this theorem can be found in the appendix.

\begin{theorem}
  If data structures A and B are lock-free, then the hybrid data structure built from A and B is also lock-free.
\end{theorem}
\begin{proof}
  See appendix \ref{proof:lf}.\rn{Need proof intuition!!}
\end{proof}

\subsection{Observable Equivalence of hybrid data structure}

In this section we prove that the hybrid data structure behaves exactly like the
original data structure, i.e., there is no way for the client to distinguish
the hybrid from the original data structures.
Specifically, we show that for each step a $\sigma^{Hybrid}$ takes, based on its schedule and its client workloads,
there exists an equivalent series of steps in a simple $B$ structure that remains observably equivalent to the hybrid.

\begin{theorem}
  If $\sigma^B \approx \sigma^{Hybrid}$ where
  $\sigma^B = (t_B,H_B,\_,M,{\Trace}_B,{\Pool}_{B})$,
  $\sigma^{Hybrid} = (t,H_{Hybrid},\_,M^+,\_,{\Pool}_{Hybrid})$ and
  ${\Pool}_{B}$ is obtained by removing all @transition@ method calls in
  ${\Pool}_{Hybrid}$, then for any scheduler $\pi$, there exists a scheduler
  $\pi'$ such that, if $\sigma^{Hybrid} \rightarrow_{\pi} \sigma^{Hybrid}_1$
  then ${\sigma^{B}} \twoheadrightarrow_{\pi'} \sigma^{B}_1$, where
  $\sigma^B_1 \approx \sigma^{Hybrid}_1$.
\end{theorem}


\new{Here, we do not need a fine-grained correspondence between intermediate
  steps of Hybrid and $B$ methods.  Rather, $\Pool_B$ contains a pristine copy
  of method $m_i$ for each method expression sharing the same label in
  $\Pool_{Hybrid}$---irrespective of whether it is already partially evaluated
  within $\Pool_{Hybrid}$.  Whereas original methods $\method{m_i}{e}$ are replaced
  by pristine $B$ versions, \il{transition} calls are instead replaced by ``No-Ops'', i.e.
  \il{return ()}.  It is only when the hybrid data structure {\em completes} a method
  that we must reduce $B$ to match the hybrid.}

\begin{proof}
  See appendix \ref{proof:eq}. \rn{Need proof intuition!!}
\end{proof}

\section{Implementation, Case Study}\label{sec:impl}

In \cref{sec:IORef}, we explain the details of our Freezable
IORef prototype in Haskell, which we use to build an example adaptive data
structure to evaluate.


\subsection{Example adaptive data structure}
\label{sec:app}

To demonstrate a concrete instantiation of the hybrid data structure, we
built a hybrid that converts a Ctrie~\cite{Prokopec:2012:CTE:2145816.2145836} to
a purely functional hashmap. We {use our implementation of} freezable @IORef@s
from \Cref{sec:IORef} for all the mutable memory locations inside the CTrie data
structure.
The idea of this pairing is to use a scalable structure that is better suited
for concurrent writes in one phase, followed by a pure data structure that uses
less memory and supports exact snapshots in the later.

\subsubsection{Pure-in-a-box HashMap}

To begin with, we have a @PureMap@ data structure, which is simply a
@HashMap@~\cite{Bagwell01idealhash} from the unordered-containers
package\footnote{https://hackage.haskell.org/package/unordered-containers}
inside a mutable reference. Because this is our {\em B} data structure, we do
not need a freezable reference. \nnew{\il{PureMap} retains lock-freedom by
  performing {\em speculative} updates to the IORef using CAS.}

\begin{snippet}
import Data.HashMap.Strict
type PureMap k v = IORef (HashMap k v)
\end{snippet}

\subsubsection{Ctrie and Cooperative Conversions}

For our scalable concurrent structure, we use an implementation from the ctrie
package\footnote{https://hackage.haskell.org/package/ctrie}. We modify the
import statement to switch the implementation to freezable
@IORef@s\footnote{This kind of substitution could be more easily done with a
  full featured module system like the forthcoming Backpack for
  Haskell~\cite{kilpatrick2014backpack}.}.
We then add a @freeze@ operation to the library. The simplest freeze recursively walks the
tree structure and freezes all the mutable cells in a depth-first order.

\mypara{Freeze+convert}

A simple optimization for a user (or a compiler) to perform is to fuse the loops
within freeze and convert, which are treated separately in our proofs, but occur
consecutively in the code. To this end, we extend Ctrie's freeze to invoke a
user-specified function as well, with which we can copy data to the B structure.

\mypara{Cooperation}
\label{sec:cooperative}

Finally, we improve this freeze+convert operation using a single traversal
function that is randomized and {\em reentrant}.  That is, multiple calls to the
same function on different threads should speed up the process, rather than
cause any conflicts.  The algorithm we implement for cooperation works for a
variety of tree datatypes.  It depends on keeping another, extra mutable bit per
@IORef@, to mark, first, when the reference is frozen, and second, when it is
{\em done} being processed, and its contents are already reflected in the B data
accumulator. In pseudocode:

\begin{snippet}[basicstyle=\small\ttfamily]
  freezeConvert tr acc = do
    p <- (*@\text{\em \textsl{my thread's random permutation}}@*)
    freezeRef (*@\text{\em \textsl{this node, }}@*)tr
    forEach (*@\text{\em \textsl{subtree {\em t}, reordered by}}@*) p:
      if (*@\text{\em \textsl{{\em t} marked done}}@*)
      then (*@\text{\em \textsl{do nothing}}@*)
      else (*@\text{\em \textsl{recursively process {\em t}}}@*)
           (*@\text{\em \textsl{union into {\em acc}}}@*)
    markDone (*@\text{\em \textsl{this node}}@*)
\end{snippet}

This algorithm freezes mutable intermediate nodes of the tree (in our case the
Ctrie) on the way down.  This means that each mutable reference is frozen before
it is read, and thus the {\em final} version of the data is read.
On the way back up, each intermediate node is marked {\em done} only after all
of its subtrees are accounted for in the output accumulator.
Yet because every thread gets its own random permutation of subtrees, multiple
threads calling @freezeConvert@ are likely to start with different subtrees.

In between freezing a subtree and marking it as done, it is possible for
redundant work to occur, so it is critical that the union operation into the
accumulator be {\em idempotent}.  But once each subtree is completely processed
{\em and} placed in the accumulator, other threads arriving at the subtree will
skip it.  This strikes the balance between lock-freedom---which requires that
any thread complete the whole job if the others are stalled---and minimization
of redundant work.

In our implementation, results are unioned into the target accumulator using
@atomicModifyIORef@ and @HashMap@'s @union@ operation.  However, processing
subtrees separately and unioning their results can yield a quadratic conversion
algorithm, in contrast to sequential conversion, which simply does a preorder
traversal, threading through the accumulator and inserting one element at a
time.  Specifically, when converting to a @HashMap@, this means calling the
@insert@ function rather than the @union@ function over @HashMap@s.

We resolve this tension by ``bottoming out'' to a preorder traversal after a
constant number of rounds of the parallel recursion.  Because we have a 64-way
fan-out at the top of a Ctrie, we in fact bottom out after only one level.


\section{Evaluation}\label{sec:eval}

In this section, we perform an empirical evaluation of the data structures
discussed in \Cref{sec:impl}. The machine we use is a single-socket Intel{\textregistered} Xeon{\textregistered}
CPU E5-2699 v3 with 64GB RAM, but for section \Cref{ssec:mapbench} we use
Intel{\textregistered} Xeon{\textregistered} CPU E5-2670 with 32GB RAM.

\setlength{\textfloatsep}{10pt}

\subsection{Heating up data}


Before evaluating our Ctrie that transforms into a PureMap, we evaluate the
reverse transition, from PureMap to Ctrie: a scenario where transition can be
triggered automatically under contention.  This is a simple case where a hybrid
data structure should be able to dynamically combine advantages of its two
constituent data structures.

We conduct a benchmark in the style of the popular Synchrobench
framework~\cite{Gramoli:2015:MYE:2688500.2688501} for assessing concurrent data
structures.  We measure the average throughput of three implementations: PureMap
(HashMap in a box mentioned in \cref{sec:app}), Ctrie and WarmupMap. Here
WarmupMap is the hybrid data structure that starts with PureMap and changes to
Ctrie on transition. The heuristic used to trigger a transition is based on
failed CAS attempts (contention). If a thread encounters 2 CAS failures
consecutively, it will initiate the transition from PureMap to Ctrie.

\begin{figure}[!htb]
  \centering
  \includegraphics[scale=0.70]{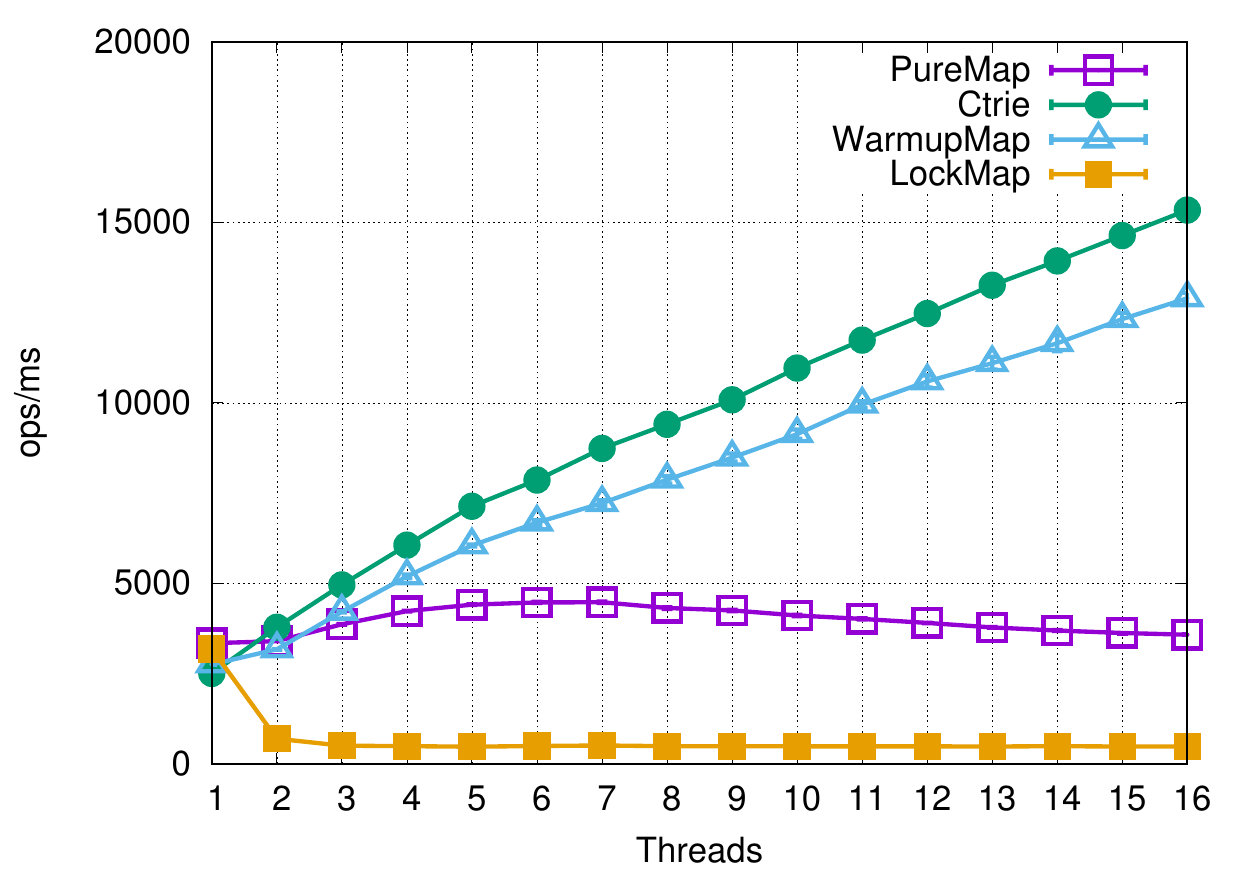}
  \caption{Heating up data. Average throughput of random operations (50\% get,
    25\% insert and 25\% delete), each run is 500ms and average over 25 runs.}
  \label{fig:coldtohot}
\end{figure}

In this benchmark, every thread randomly calls @get@, @insert@ and @delete@ for 500ms. The probability distribution of
operations are: 50\% @get@, 25\% @insert@ and 25\% @delete@. We test for 1 to 16 threads, and measure the average
throughput (ops/ms) in 25 runs. The results are shown in \Cref{fig:coldtohot}.

As we can see from the figure, the PureMap performs best on one thread but scalability is very poor;
Ctrie has good scalability;
and WarmupMap has better performance than Ctrie in one thread since it remains in the PureMap state (no CAS failures
on one thread). Conversely, the adaptive version transitions to Ctrie so it has much better scalability than PureMap.
The gap between Ctrie and WarmupMap is due to the cost of transitioning plus the
cost of extra indirections in freezable references.
As a baseline, we also include a version using the same HashMap but with a
coarse-grained lock around the data structure (an MVar in Haskell).

\subsection{Parallel freeze+convert}

Next, we check that the algorithm described in \Cref{sec:cooperative} improves
the performance of the freeze+convert phase by itself. For example, on 1 thread,
the simple preorder traversal takes 3.37s to freeze and convert a 10 million
element Ctrie, and the randomized algorithm matches its performance (3.36s).
But on 16 threads, the randomized version speeds up to 0.41s for the same freeze
plus conversion\footnote{Unintuitively, the sequential version also speeds up
  with more threads, because it utilizes parallel garbage collection and
  conversion is GC intensive.  It reaches a peak performance in this example of
  1.71s.}.


\subsection{Cooling down data}

In this benchmark, we compare the performance of {\em AdaptiveMap} (Ctrie to Pure)
against
PureMap and Ctrie.
No heuristic is used here, the programmer calls the @transition@ method
manually based on known or predicted shifts in workload.

\begin{figure}[!htb]
  \centering
  \includegraphics[scale=0.70]{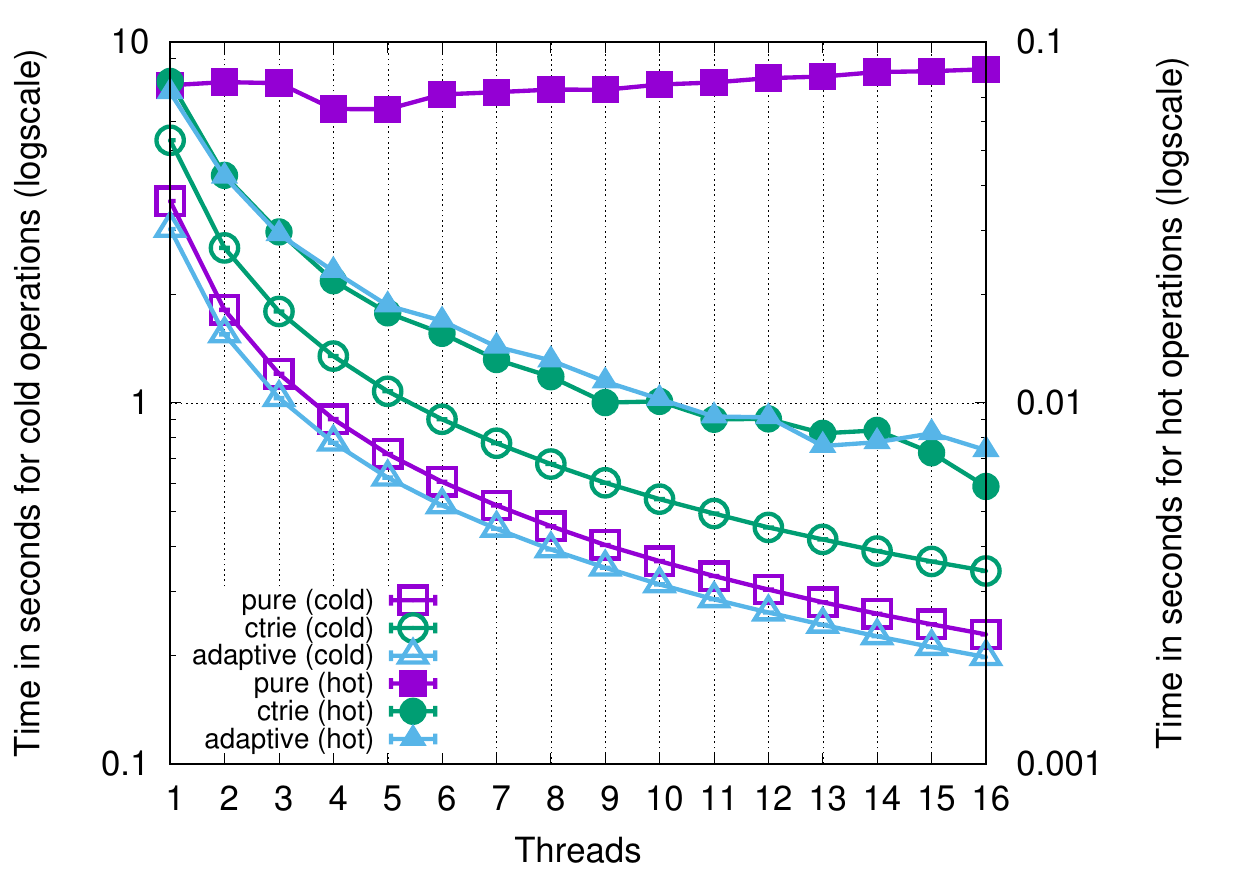}
  \caption{Median time (logscale) over 25 runs for $10^5$ hot operations and
    $2 \times 10^7$ cold operations, measured separately, with initial size 0,
    and randomly generated keys in the range $[0, 2^{32}]$. Solid dots
    correspond to the right y-axis, hollow dots correspond to the left y-axis.}
  \label{fig:hot-and-cold}
\end{figure}

We measure the time to complete a fixed number of @insert@ operations (hot
phase) or @get@ operations (cold phase), distributed equally over varying number
of threads. In \Cref{fig:hot-and-cold}, we break these phases down individually,
for the AdaptiveMap, running the hot operations when it is in the Ctrie state,
and cold operations after it transitions to the PureMap state.

We can see that AdaptiveMap closely tracks the performance of Ctrie in the hot
phase, and PureMap in the cold phase.  Again, the gap between Ctrie and
AdaptiveMap shows the overhead of extra indirections in freezable references.
\Cref{fig:hot-and-cold} is representative of what happens if the adaptive data
structure is put into the correct state {\em before} a workload arrives.

\begin{figure}[!htb]
  \centering
  \includegraphics[scale=0.70]{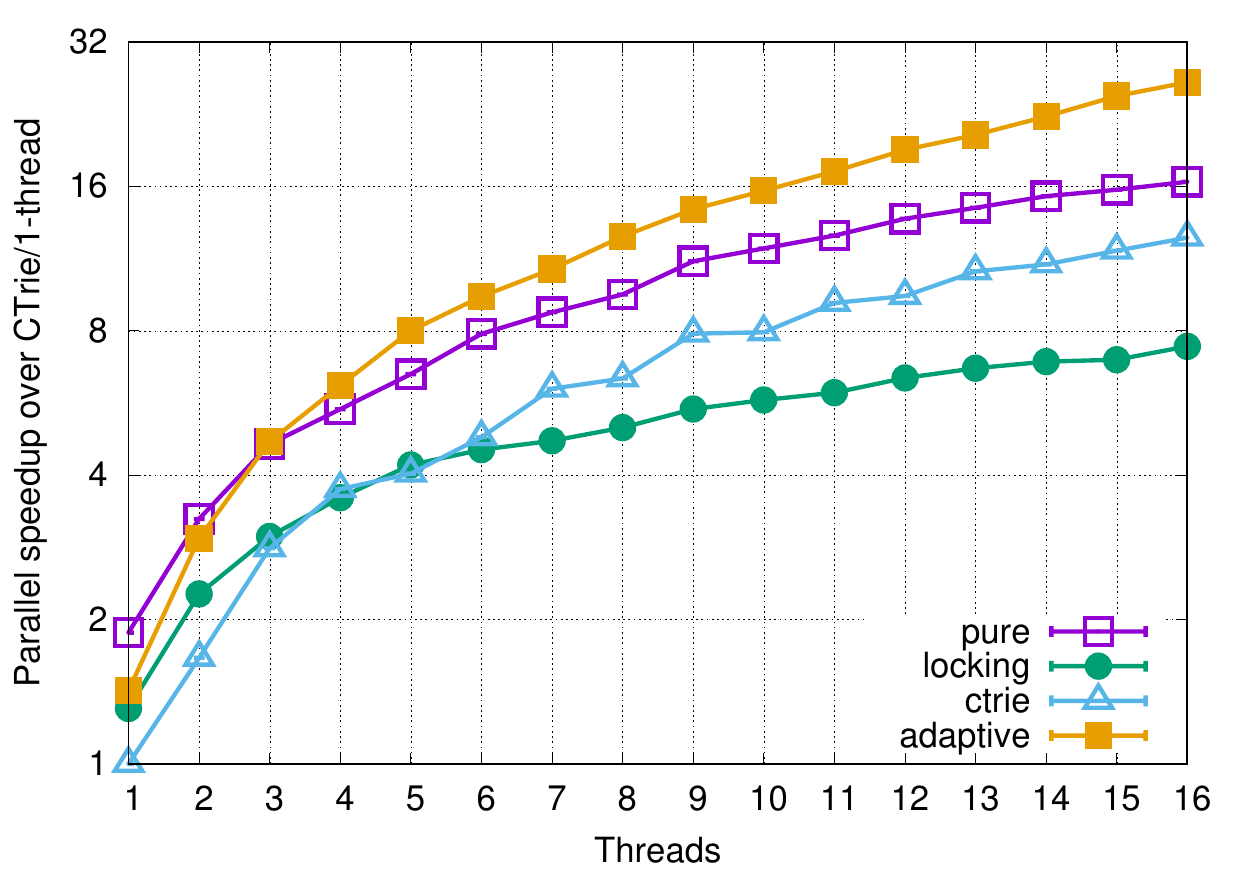}
  \caption{Parallel speedup relative to time taken by Ctrie on a single thread,
    median over 25 runs, for $10^5$ hot operations, followed by transition, and
    then $2 \times 10^7$ cold operations, with initial size 0, and randomly
    generated keys in the range $[0, 2^{32}]$}
  \label{fig:hotcold}
\end{figure}

In the next benchmark, ``hotcold'', we measure the total time to complete a
fixed number of operations, divided in a fixed ratio between @insert@ (hot
phase) and @get@ (cold phase), with the adaptive data structure
transitioning---during the measured runtime---from Ctrie to PureMap, at the
point where the hot phase turns to cold. The results are shown in
\Cref{fig:hotcold}.

Here the latency of the transition starts out at 0.6ms on a single
thread, and reduces to 0.2ms on 16 threads, because more threads can accomplish
the freeze/convert faster using the randomized algorithm of \Cref{sec:impl}.

In this two-phase scenario, because both the non-adaptive variants have to spend
time in their ``mismatched'' workloads, the adaptive algorithm comes out best
overall.
One interesting outcome is that Ctrie is bested by PureMap in this case.  In a
scenario of concurrent-reads rather than concurrent-writes, the purely
functional, immutable data structure occupies less space and can offer higher
read throughput in the cold phase.

\begin{figure}
  \centering
  \includegraphics[scale=0.70]{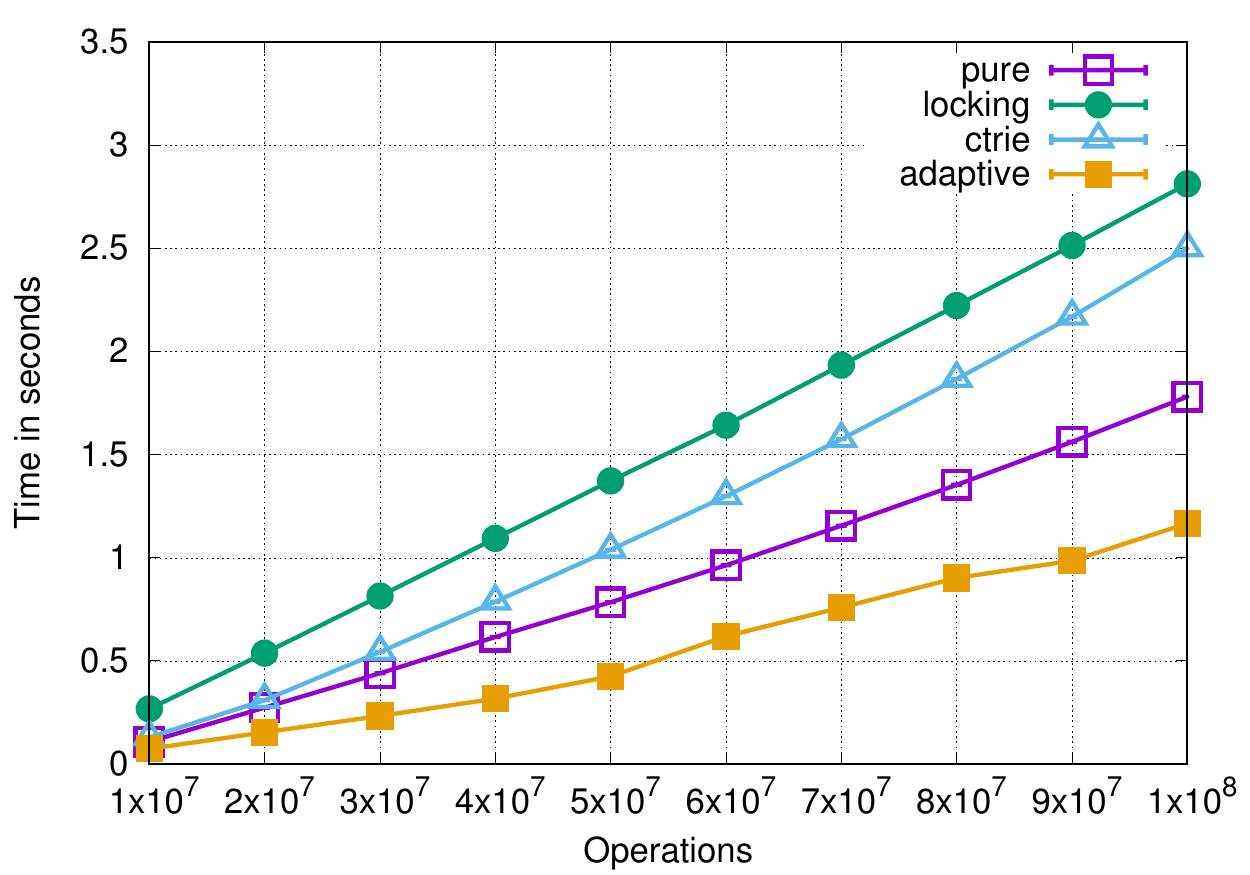}
  \caption{Median time over 25 runs for hot operations followed by transition then cold operations, with hot-to-cold
    ratio of $1:200$, equally divided over 16 threads, with initial size 0, and randomly generated keys in the range
    $[0, 2^{32}]$}
  \label{fig:hotcold-size}
\end{figure}

Next, we scale the number of operations (and maximum data structure size) rather
than the threads.  We measure the total time including transition time for a
mixed workload of @insert@ (hot) operations followed by transition, then @get@
(cold) operations, keeping the hot-to-cold ratio constant, and varying over the
total number of operations, divided equally over a fixed number of threads. The
results in \Cref{fig:hotcold-size} show that AdaptiveMap beats both PureMap and
Ctrie on such a workload.

At the right edge of \Cref{fig:hotcold-size}, the data structure grows to
contain 500K elements, and transition times are 50ms, which is equivalent to a
throughput of 10 million elements-per-second in the freeze and convert step.

\subsection{Compacting data}

As an additional scenario, we use our technique to build an adaptive data
structure that transitions from a PureMap to a CompactMap, which is a PureMap
inside a {\em compact} region~\cite{cnf-icfp15}. Objects inside a compact region
are in normal form, immutable, and do not need to be traced by the garbage
collector. This is an excellent match for cooling down data; we build a large
PureMap (hot phase), and transition to a CompactMap for read operations (cold
phase).

We measure total number of bytes copied during garbage collection, for
CompactMap, comparing against ``No-op'' and PureMap.
No-Op is no data structure at all, consisting of empty methods.
GC is triggered by simulating a
client workload, generating random keys using a pure random number generator and
looking up values from the map. The map is filled by running insert operations
on random keys, followed by transition, and then equally many get
operations. The workload is equally divided among 16 worker threads.

\begin{figure}
  \centering
  \includegraphics[scale=0.70]{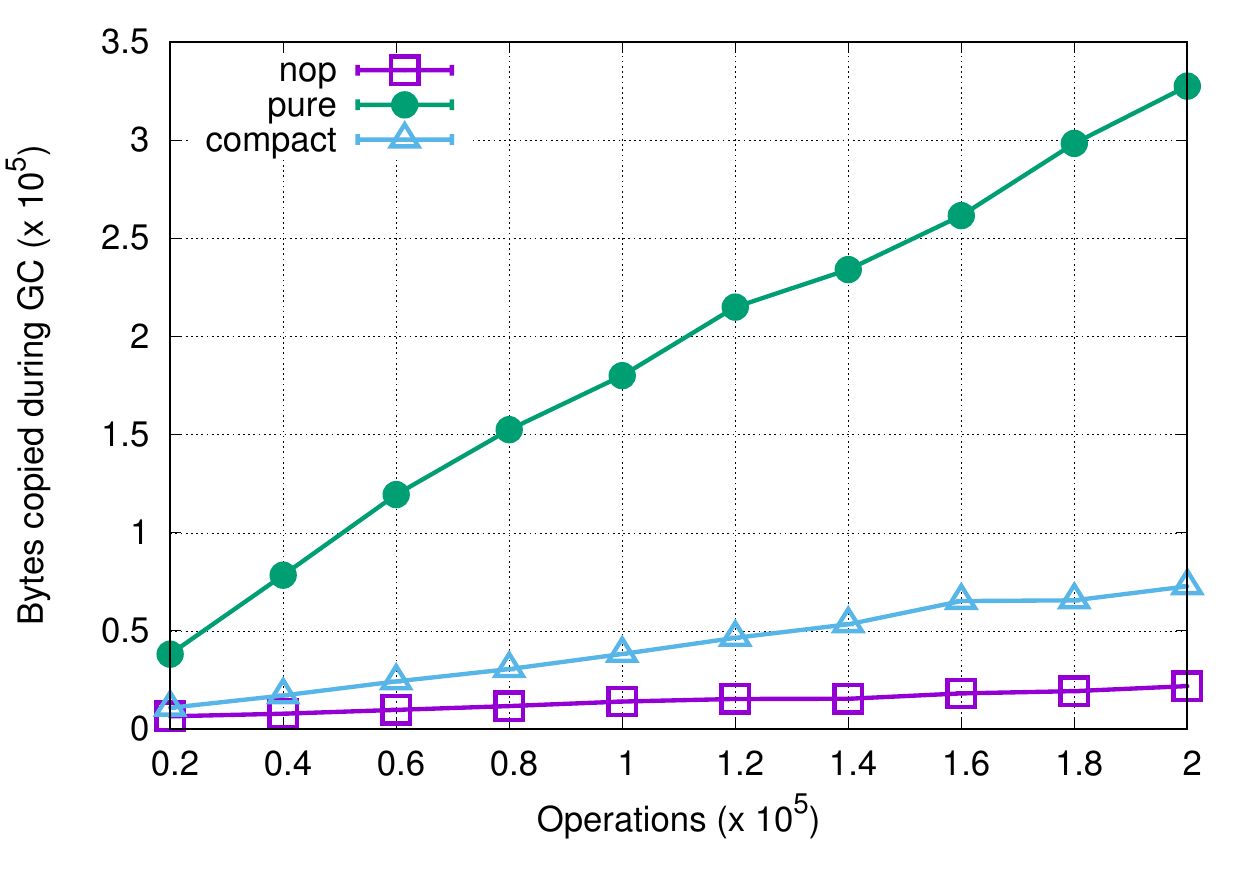}
  \caption{Bytes copied during GC against varying number of operations,
    averaging over 25 iterations, with hot-to-cold ratio of $1:1$, equally
    divided over 16 threads, with initial size 0, and randomly generated keys in
    the range $[0, 2^{63}]$}
  \label{fig:cold-gc}
\end{figure}

The results are shown in \Cref{fig:cold-gc}.  As the number of operations
increases, the version that adapts into a compact form has fewer bytes to copy
during GC.  Even the No-Op version copies {\em some} amount, as it runs the
client workload (including RNG).

Compacting data provides one example of dealing with a read-heavy phase.  Other
examples include creating additional caching structures or reorganizing data.

\subsection{Shifting hot spots}
\label{ssec:mapbench}

In this benchmark we consider an idealized document server where one document is
hot and others are cold, and these roles change over time.
We have 2000 @CTrie Int ByteString@ (each representing a dynamic document, in
chunks), for phase $i$ ($i \in [1,2000]$) every thread has a 50\% chance to
perform a hot operation on $i$-th map, otherwise it chooses a map uniformly at
random and performs a (cold) operation.  For a hot operation, it is 80\% insert
and 20\% lookup.  For a cold operation, it is 80\% lookup and 20\% insert.\par

\begin{figure}
  \centering
  \includegraphics[scale=0.70]{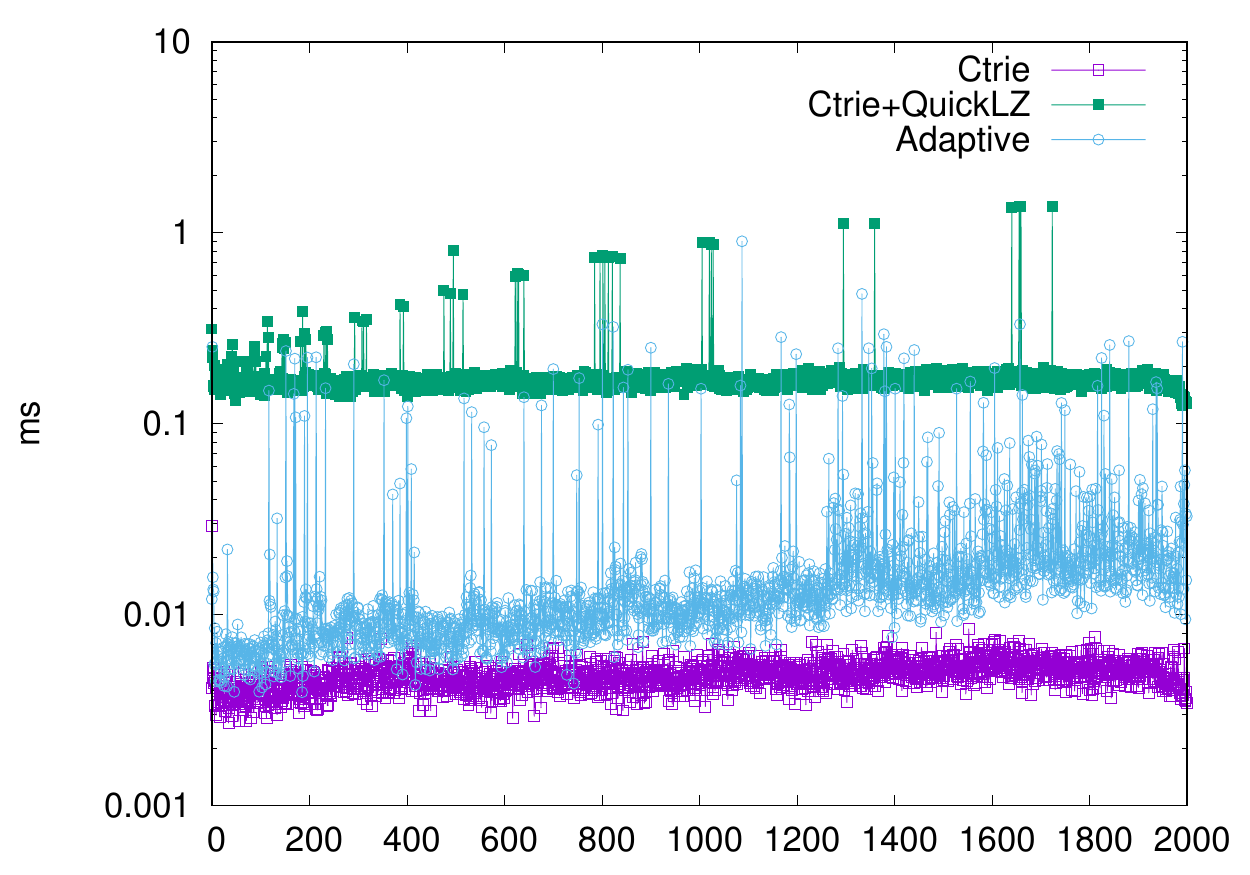}
  \caption{The average latency for each round of 200 operations; y-axis is time in logscale.}
  \label{fig:mapbench}
\end{figure}

We consider three implementations: first, plain Ctrie; second, Ctrie with {\em
  QuickLZ} to compress the @ByteString@ (and decompress on all lookups), and,
third, {\em adaptive map} which will transition from Ctrie to Ctrie+QuickLZ
after each phase, when a document switches from hot to cold.  In this benchmark
we use 8 threads---the average latency for a method call is in figure
\Cref{fig:mapbench} and the maximum memory residency for the whole run is shown
in \Cref{fig:peak-mem}.
According to these results, Ctrie+QuickLZ is significantly slower but saves
memory, whereas adaptive map achieves a balance between the two implementations:
saving memory without introducing too much overhead.

\begin{figure}
\begin{center}
  \begin{tabular}{ | c | c | c | }
    \hline
    Ctrie & Ctrie+QuickLZ & Adaptive \\ \hline
    16240 megabytes & 7738 megabytes & 13194 megabytes \\ \hline
  \end{tabular}
\end{center}
  \caption{Peak memory usage on server variants.}
  \label{fig:peak-mem}
\end{figure}

\section{Conclusion}\label{sec:con}

In this paper, we proposed a simple technique that requires a minimum amount of
work to add an adaptive capability to existing
lock-free data structures.
%
This provides an easy starting point for a programmer to
build adaptive lock-free data structures by composition.
The hybrid data structure can be further optimized to better serve
practical client workloads, as the programmer sees fit.

Freezable references emerged as a useful primitive in this work.
In the future, we plan
to work on a transition method that preserves lock-freedom without sacrificing
performance, and on techniques to automatically detect, or predict, when
transition is needed.








\appendix
\section*{Appendix}

\section{CAS in a purely functional language}\label{sec:tickets}


\new{Our formal treatment in \Cref{sec:correctness} benefits from using a purely
  functional term language which separates effects into a monad. This means,
  however, that allocation is not a side effect and that pure values lack
  meaningful pointer identity.}

\new{How then to implement compare-and-swap?  The trick that Haskell uses is to give values
  identity {\em by virtue of being pointed to by a mutable reference}.  As soon as
a value is placed in a reference with \il{writeIORef}, it becomes the subject of
possible CAS operations.  We can observe a kind of {\em version} of the value by
reading the reference and getting back an abstract ``ticket'' that encapsulates
that observation:}

\begin{snippet}
  readForCAS :: IORef a -> IO (Ticket a)
  peekTicket :: Ticket a -> a
\end{snippet}

\new{So while we may not be able to distinguish whether one value of type \il{a}
  has the same ``pointer identity'' as another, we can distinguish whether the
  value in the IORef has changed since a ticket was read.  And our CAS
  operation then uses a ticket in lieu of the ``old'' value in machine-level CAS:}

\begin{snippet}
  casIORef   :: IORef a -> Ticket a -> a
             -> IO (Bool, Ticket a)
\end{snippet}

\new{Unlike machine-level CAS, this operation returns {\em two} values: first, a
  flag indicating whether the operation succeeded, and, second, a ticket to
  enable future operations.  In a successful CAS, this is the ticket granted to
  the ``new'' value written to the memory cell.}

\subsection{Physical Equality and Double Indirection} \label{sec:double-indirection}

The above semantics for CAS depends on a notion of value equality, e.g., where
we test $H(\loc) = old$.
However, atomic memory operations in real machines work on single words of
memory.  Thus, {\em physical, pointer equality} is the natural choice for a
language incorporating CAS.  \new{But, as described in \Cref{sec:tickets},} this
is at odds with a purely functional language which does not have a built-in
notion of pointer identity or physical equality.  (This is also reflected in the
fact that the separate, pure, term reduction (\Cref{fig:semantics1}) makes
no use of the heap, $H$, \new{and does not mention locations $\loc$}.)

How do we square this circle?  The basic solution is to introduce a
mechanism to selectively create a meaningful pointer identity for values which
we wish to store in atomically-updated references.
\new{The concept of a ``Ticket'' shown in \Cref{sec:tickets} is the the}
approach used by the Haskell \il{atomic-primops} library that exposes CAS to
users.  It uses a GHC-specific mechanism for establishing meaningful pointer
equality and preventing compiler optimizations (such as unboxing and
re-boxing) that would change it.

In our simple formal language we instead restrict CAS to operate on references
containing {\em scalars}, values representable as a machine word.  In
particular, we consider both locations $\loc$ and integers $n$ to be scalar values.
Accordingly, we use a restricted typing rule for \cas{}, shown in
\Cref{fig:types}.


Because we can perform CAS on locations, we can enable a similar approach as the
\il{atomic-primops} library using {\em double indirection}.  That is, because
references already correspond to heap locations, simply storing a value
$v$ in a reference, allows us to use its location $\loc$ in lieu of $v$.
%
For instance, $\Ref (\Int,\Int)$ would not support atomic modification,
whereas $\Ref (\Ref (\Int,\Int))$ would allow us to achieve the desired effect.

To follow this protocol, pure values to be used with CAS must be ``promoted'' to
references with \newrefName{} (references which are never subsequently modified).
As part of our implicit desugaring to our core language, we treat all CAS
operations on non-scalar types as implicitly executing a \newrefName{} first.
For instance, the root of the hybrid data structure must be treated thus,
because it stores compound data values such as @(AB _ _)@.
In the operational semantics of the next section, this means that CAS operations
on such types consume two time steps instead of one, which is immaterial to
establishing upper bounds for progress (lock-freedom).

\section{Freezable References: Implementation}\label{sec:IORef}


\Cref{fig:IORef} gives a Haskell definition for a freezable reference as a
wrapped @IORef@. It is nothing but an @IORef@ plus one extra bit of information,
indicating frozen status.
We also define an exception @CASFrznExn@, which is
raised when a thread attempts a CAS operation on a frozen @IORef@.
This implementation adds an extra level of indirection, but as a result it can
be atomically modified {\em or} atomically frozen by changing one physical
memory location.  \new{An implementation with compiler support might, for
  example, use pointer tagging to store the frozen bit.}

\Cref{fig:freezeIORef} shows our @freezeIORef@ function. Once this succeeds, the
reference is marked as frozen and no further modification is allowed. Any
attempt to modify a frozen @IORef@ will raise the exception @CASFrznExn@.

We wrap and
re-export relevant functions in @Data.IORef@, such as @writeIORef@ and @atomicModifyIORef@,
so that a freezable @IORef@ can be
used without any code modifications---only a change of module imported. We also
provide all functions related to @IORef@ in
@Data.Atomics@\footnote{https://hackage.haskell.org/package/atomic-primops},
such as @casIORef@.
%
%
The implementation
of these functions is straightforward; they are easily written by pattern
matching on the wrapper @IOVal@.  Read operations ignore the frozen bit, but
write operations throw an exception when applied to a frozen reference.
\if{1}
     For instance:

     \begin{snippet}
     import Control.Exception
     import Data.Atomics as A

     casIORef :: IORef a
              -> Ticket (IOVal a)
              -> a
              -> IO (Bool, Ticket (IOVal a))
     casIORef (IORef ref) tik a = do
       case A.peekTicket tik of
         Frozen _ -> throwIO CASFrznExn
         Val    _ -> A.casIORef ref tik (Val a)
     \end{snippet}
\fi{}

\begin{figure}[t]
\noindent
\begin{minipage}{0.48\textwidth}
\begin{snippet}
module Data.IORef.Freezable (...) where

import qualified Data.IORef as IR
import Control.Monad.Except

newtype IORef t = IORef (IR.IORef (IOVal t))
data IOVal t = Val t | Frozen t

data CASFrznExn = CASFrznExn
deriving (Show, Exception)
  \end{snippet}
  \caption{Definition of the freezable variant of IORef.}
  \label{fig:IORef}
  \label{fig:exception}
\end{minipage}
\begin{minipage}{0.48\textwidth}
\begin{code}
freezeIORef :: IORef a -> IO ()
freezeIORef r = loop r =<< readForCAS r
  where
  loop ref tik = do
   (flag, tik') <- tryFreezeIORef ref tik
   unless flag (loop ref tik')

tryFreezeIORef :: IORef a
            -> Ticket (IOVal a)
            -> IO (Bool, Ticket (IOVal a))
tryFreezeIORef (IORef ref) tik = do
  case peekTicket tik of
    Frozen _ -> return (True, tik) (*@\label{l_frozen}@*)
    Val    a -> A.casIORef ref tik (Frozen a)
  \end{code}
  \caption{Freeze or attempt to freeze a single reference}
  \label{fig:freezeIORef}
\end{minipage}
\end{figure}



\section{Correctness Proof}

\begin{theorem}
  If data structures A and B are lock-free, then the hybrid data structure built from A and B is also lock-free.
  \label{proof:lf}
\end{theorem}
\begin{proof}
  For any configuration of the hybrid data structure, $\sigma^{Hybrid}\allowbreak=(t,H,F,M^+,\Trace,\Pool)$, and schedule $\pi$,
  there exists an upper bound function $f$ such
  that, some thread $\tau_j \in \threadids{}$ will finish executing method $m_i \in M^+$ in $f(\sigma^{Hybrid})$
  timesteps.

  {We use shorthands $\sigma^A$ and $\sigma^B$ to refer to {\em projections} of
    $\sigma^{Hybrid}$ that take only the subset of the state relevant to the A
    or B structures, i.e. viewing $\sigma^{Hybrid}$ {\em as though it were an A}.}
  Likewise, $H_A$ and $H_B$ refer to projections of the heap, $H_{Hybrid}$.

  Further, to account for the extra read (i.e. CAS) at the beginning of each hybrid method
  on line \ref{l_rw_read}, we abbreviate $f_A'(x) = f_A(x) + |\threadids|$ to
  refer to the updated bound on A (and likewise for $f_B'$).  The reason the increase is proportional to
  the number of threads, is that an adversarial scheduler can cause each thread
  to spend one operation here before getting to any actual calls to A's methods.

  To define $f(\sigma^{Hybrid})$, here we reason by cases over the current state, $\sigma^{Hybrid}$.
  There are three cases to consider based on the current state of the heap, $H_{Hybrid}$:

  \begin{description}
    [style=nextline,font=\normalfont\textbullet\space,leftmargin=*,listparindent=\parindent,topsep=1em,itemsep=1em]
  \item[{\bf A state:} {$H(H(\loc_{hy}))$ is an $A$ value}
  ]

    For methods other than transition, by the lock-freedom of data structure A,
    there must be some method which finishes execution in $f_A'(|H_A|)$
    timesteps.
    This bound, representing A's lock-freedom, holds if there is no thread
    which executes transition and passes line \ref{l_tran_cas} on or before time
    $t + f_A'(|H_A|) + 2$.  \new{Here the additional ``$+ 2$'' addresses the
      fact that the original method may throw an exception on the very last time
      step during which it would have succeeded, and then require two more time
      steps to cross line \ref{l_tran_cas}.}


    If a thread passes line \ref{l_tran_cas} in time $t'$, such that
    $t' \leq t + f_A'(|H_A|) + 2$, and the configuration at that time is
    $\sigma^{Hybrid}_{t'} = (t', H', F, M^+, \allowbreak \Trace ', \Pool ')$
    where {$H(H(\loc_{hy}))$ is an $AB$ value}, then \new{the time
    for a transition or another method to subsequently complete is bounded by}
    $f_{tran}(|H_{Hybrid}|+t'-t)$---where $f_{tran}$
    will be defined below in the next case.  So, we have
    $f(|H_{Hybrid}|) = f_A'(|H_A|) + 2 + f_{tran}(|H_{Hybrid}| + t' - t)$.



  \item[{\bf AB state:} {$H(H(\loc_{hy}))$ is an $AB$ value} ]


    We reach this state only if some thread executing transition already
    passed line \ref{l_tran_cas}, and none of the threads executing transition
    pass the linearization points at lines
    \ref{l_tran_casB1} or \ref{l_tran_casB2}. In this state, we have $K \ge 1$ threads executing transition and $K' \ge
    0$ threads executing other methods in $\Pool{}$.

    If the scheduler only schedules the $K'$ threads executing methods other
    than transition --- \new{never entering line \ref{l_rw_trans}} --- then some
    method will finish execution in $f_A'(|H_A|)$ timesteps.
%
%
    Otherwise, at least some thread takes a step inside transition.  Examining
    \Cref{fig:rw_op_lf}, if thread $\tau_j$ tries to $cas(\loc,\_,\_)$ where
    $\loc \in F$, it will raise an exception, and according to line
    \ref{l_rw_trans2}, this causes $\tau_j$ to start executing transition, so
    the next time it gets scheduled, the configuration of $\sigma^A$ in the
    future will behave as if the scheduler never scheduled $\tau_j$. Any {\em
      new} method call entering the picture, according to line \ref{l_rw_trans},
    will just call transition.  Since there are only $|\threadids{}|$ threads,
    after $|\threadids{}| \times f_A'(|H_A|)$ timesteps, either every thread is
    executing transition, or some method other than transition completes. For
    each timestep, at most one new memory cell can be allocated, so if no method
    completes, the heap size of $\sigma^A$ is at most $|H_A| + |\threadids{}|
    \times f_A'(|H_A|)$.

    After the first thread does a successful CAS on line \ref{l_tran_cas}, the
    address $H(\loc_{hy})$ will not change until some thread is on line
    \ref{l_tran_casB1} or \ref{l_tran_casB2}, and tries to CAS
    $H(\loc_{hy})$. The first thread that executes this CAS must succeed because
    $\loc_{hy} \notin F$, and the old value is the current value in the
    heap. Since the runtimes of @freeze@ and @convert@ only depend on the heap
    size, some thread must finish executing @freeze@
    \new{(lines \ref{l_freeze1} or \ref{l_freeze2}) in
    $|\threadids{}| \times (f_{frz}(|H_A| + |\threadids{}| \times f_A'(|H_A|)))$ timesteps,
    and then \il{convert} (lines \ref{l_convert1} or \ref{l_convert2}) in
     $|\threadids{}| \times (f_{conv}(|H_A| + |\threadids{}| \times f_A'(|H_A|))) + 1$
    timesteps.  Finally threads must then must attempt the CAS in lines
    \ref{l_tran_casB1} or \ref{l_tran_casB2}.}
    \new{After all freezes and conversions have completed, a thread can only
      fail at these lines if another succeeds, and so in the very next timestep
    one thread must succeed in changing the state to B.  This last CAS is the
    linearization point and completes the transition.}
    So after
    \begin{align*}
      f_{tran}(|H_{Hybrid}|) &= |\threadids{}| \times (f_{frz}(|H_A| + |\threadids{}| \times f_A'(|H_A|))) \\
                           &+ |\threadids{}| \times (f_{conv}(|H_A| + |\threadids{}| \times f_A'(|H_A|))) \\
                           &+ |\threadids{}| \times f_A'(|H_A|) + 1
    \end{align*}
    timesteps, either a data structure method completes or some thread finishes
    the transition method.

  \item[{\bf B state:} {$H(H(\loc_{hy}))$ is a $B$ value} ]

    If a thread calls transition at this state, execution passes through
    line \ref{l_tranB} and the transition finishes executing immediately.  Further
    completing this method constitutes progress vis-\`a-vis lock freedom.


    Other new method calls execute line \ref{l_rw_commit_B}, thus calling the corresponding
    method from $\sigma_B$.
    %
    By lock-freedom on data structure B, there must be some method that finishes
    execution in $f_B'(|H_B|)$ timesteps.  But some time may be wasted on
    threads that have not yet finished executing transition.  Being in the B
    state means that at least {\em one} thread has completed a transition, but
    there may be others ongoing.
    These ongoing transitions will finish in $f_{tran}(|H_{Hybrid}|)$ timesteps.

    For those methods executing inside the B code (line \ref{l_rw_commit_B})
    we have the bound from B, $f_B'(|H_B|)$.  Methods executing the A code (line \ref{l_rw_commit_A})
    raise
    an exception in $f_A'(|H_A|)$ steps when they try to commit with a CAS (or
    earlier).  Methods which follow that exception path, according to line
    \ref{l_rw_trans3}, fall back under the B case after wasting some steps to
    get there.
    So we have, $f_{Hybrid}(|H_{Hybrid}|) = f_A'(|H_A|) +
    f_{tran}(|H_{Hybrid}|) + f_B'(|H_B|)$.
  \end{description}

  We defined the upper bound function $f(\sigma^{Hybrid})$ for these three
  cases, which concludes the proof.
\end{proof}

\begin{theorem}
  If $\sigma^B \approx \sigma^{Hybrid}$ where
  $\sigma^B = (t_B,H_B,\_,M,\allowbreak {\Trace}_B,{\Pool}_{B})$,
  $\sigma^{Hybrid} = (t,H_{Hybrid},\_,M^+,\_,{\Pool}_{Hybrid})$ and
  ${\Pool}_{B}$ is obtained by removing all @transition@ method calls in
  ${\Pool}_{Hybrid}$, then for any scheduler $\pi$, there exists a scheduler
  $\pi'$ such that, if $\sigma^{Hybrid} \rightarrow_{\pi} \sigma^{Hybrid}_1$
  then ${\sigma^{B}} \twoheadrightarrow_{\pi'} \sigma^{B}_1$, where
  $\sigma^B_1 \approx \sigma^{Hybrid}_1$.
  \label{proof:eq}
\end{theorem}


\new{Here, we do not need a fine-grained correspondence between intermediate
  steps of Hybrid and $B$ methods.  Rather, $\Pool_B$ contains a pristine copy
  of method $m_i$ for each method expression sharing the same label in
  $\Pool_{Hybrid}$---irrespective of whether it is already partially evaluated
  within $\Pool_{Hybrid}$.  Whereas original methods $\method{m_i}{e}$ are replaced
  by pristine $B$ versions, \il{transition} calls are instead replaced by ``No-Ops'', i.e.
  \il{return ()}.  It is only when the hybrid data structure {\em completes} a method
  that we must reduce $B$ to match the hybrid.}

\begin{proof}
  Assume $\pi(t) = \tau$, and
  $\sigma^{Hybrid} \rightarrow_{\pi} \sigma^{Hybrid}_1$ where
  $\sigma^{Hybrid}_1 \allowbreak= (t+1,H_{Hybrid_1},\_,M^+,\_)$.  We proceed by analyzing
  the four cases according to \Cref{fig:semantics2}.
  \begin{itemize}
  \item \textsc{NEW}: Let $\pi' = \pi$ and $\sigma^{B}_1 = \sigma^B$. Since this
    is not a linearization point, we have $\sigma^B \approx \sigma^{Hybrid}_1$
    and ${\sigma^{B}} \rightarrow_{\pi}^0 \sigma^{B}$.
  \item \textsc{METHODFINISH($m$)}:
    \begin{itemize}
    \item $m \equiv $ @transition@: Same as the \textsc{NEW} case, by the
      property of @conversion@, $\sigma^{Hybrid} \approx \sigma^{Hybrid}_1$.
    \item $m \equiv m_i$:
      Let
      \[
        \pi'(t) =
        \begin{cases}
          \tau & \text{for } t_B \le t \le t'' \\
          \pi(t) & \text{otherwise}
        \end{cases}
      \]
      Then,
      \begin{align*}
        {\sigma^B} & \to_{\pi'} \sigma^{B}_0 = (t_B + 1,{H_B}',F_B',M,{\Trace}_{B},\Pool_{B_0}) \\
                   & \twoheadrightarrow_{\pi'} {\sigma^{B}_0}' = (t'' - 1,{H_B}'',F_B'',M,{\Trace}_{B},\Pool_{B_0'}) \\
                   & \to_{\pi'} \sigma^{B}_1 = (t'',{H_B}''',F_B''',M,{\Trace}_{B} \doubleplus [(m_i , \_)],\Pool_{B_0''})
      \end{align*}
      So, $\pi'(t)$ initiates sequential execution of $m_i$. Since
      $\sigma^B \approx \sigma^{Hybrid}$, we have
      $\sigma^B_1 \approx \sigma^{Hybrid}_1$.
    \end{itemize}
  \item \textsc{CAS}: Same as the \textsc{NEW} case. Note that if {$cas$
    fails and raises exception at line \ref{l_rw_trans2}, we know that $\tau_j$
    will call $m_i$} once it finishes executing @transition@.  \new{But this is
    an internal matter; because this recursion happens inside a top-level
    $\method{m_i}{}$ form, the corresponding $m_i^B$ method does not update.}
  \item \textsc{FRZ}: Same as the \textsc{NEW} case.
  \item \textsc{METHODPURE}: Same as the \textsc{NEW} case.
  \end{itemize}
\end{proof}

\begin{acks}
  This work was supported by a National Science Foundation (NSF) award \#1453508.
\end{acks}

{
\bibliographystyle{ACM-Reference-Format}
\bibliography{misc_bibliography/refs}


\begin{thebibliography}{00}


\ifx \showCODEN    \undefined \def \showCODEN     #1{\unskip}     \fi
\ifx \showDOI      \undefined \def \showDOI       #1{{\tt DOI:}\penalty0{#1}\ }
  \fi
\ifx \showISBNx    \undefined \def \showISBNx     #1{\unskip}     \fi
\ifx \showISBNxiii \undefined \def \showISBNxiii  #1{\unskip}     \fi
\ifx \showISSN     \undefined \def \showISSN      #1{\unskip}     \fi
\ifx \showLCCN     \undefined \def \showLCCN      #1{\unskip}     \fi
\ifx \shownote     \undefined \def \shownote      #1{#1}          \fi
\ifx \showarticletitle \undefined \def \showarticletitle #1{#1}   \fi
\ifx \showURL      \undefined \def \showURL       #1{#1}          \fi
\providecommand\bibfield[2]{#2}
\providecommand\bibinfo[2]{#2}
\providecommand\natexlab[1]{#1}
\providecommand\showeprint[2][]{arXiv:#2}

\bibitem[\protect\citeauthoryear{Afek, Attiya, Dolev, Gafni, Merritt, and
  Shavit}{Afek et~al\mbox{.}}{1993}]%
        {Afek:1993:ASS:153724.153741}
\bibfield{author}{\bibinfo{person}{Yehuda Afek}, \bibinfo{person}{Hagit
  Attiya}, \bibinfo{person}{Danny Dolev}, \bibinfo{person}{Eli Gafni},
  \bibinfo{person}{Michael Merritt}, {and} \bibinfo{person}{Nir Shavit}.}
  \bibinfo{year}{1993}\natexlab{}.
\newblock \showarticletitle{Atomic Snapshots of Shared Memory}.
\newblock \bibinfo{journal}{{\em J. ACM\/}} \bibinfo{volume}{40},
  \bibinfo{number}{4} (\bibinfo{date}{Sept.} \bibinfo{year}{1993}),
  \bibinfo{pages}{873--890}.
\newblock
\showISSN{0004-5411}
\showDOI{%
\url{http://dx.doi.org/10.1145/153724.153741}}


\bibitem[\protect\citeauthoryear{Agrawal, Leiserson, and Sukha}{Agrawal
  et~al\mbox{.}}{2010}]%
        {Agrawal:2010:HLF:1693453.1693487}
\bibfield{author}{\bibinfo{person}{Kunal Agrawal}, \bibinfo{person}{Charles~E.
  Leiserson}, {and} \bibinfo{person}{Jim Sukha}.}
  \bibinfo{year}{2010}\natexlab{}.
\newblock \showarticletitle{Helper Locks for Fork-join Parallel Programming}.
  In \bibinfo{booktitle}{{\em Proceedings of the 15th ACM SIGPLAN Symposium on
  Principles and Practice of Parallel Programming}} {\em
  (\bibinfo{series}{PPoPP '10})}. \bibinfo{publisher}{ACM},
  \bibinfo{address}{New York, NY, USA}, \bibinfo{pages}{245--256}.
\newblock
\showISBNx{978-1-60558-877-3}
\showDOI{%
\url{http://dx.doi.org/10.1145/1693453.1693487}}


\bibitem[\protect\citeauthoryear{Bagwell}{Bagwell}{2001}]%
        {Bagwell01idealhash}
\bibfield{author}{\bibinfo{person}{Phil Bagwell}.}
  \bibinfo{year}{2001}\natexlab{}.
\newblock \showarticletitle{Ideal Hash Trees}.
\newblock \bibinfo{journal}{{\em Es Grands Champs\/}}  \bibinfo{volume}{1195}
  (\bibinfo{year}{2001}).
\newblock


\bibitem[\protect\citeauthoryear{Bauman, Bolz, Hirschfeld, Kirilichev, Pape,
  Siek, and Tobin-Hochstadt}{Bauman et~al\mbox{.}}{2015}]%
        {Bauman:2015:PTJ:2784731.2784740}
\bibfield{author}{\bibinfo{person}{Spenser Bauman},
  \bibinfo{person}{Carl~Friedrich Bolz}, \bibinfo{person}{Robert Hirschfeld},
  \bibinfo{person}{Vasily Kirilichev}, \bibinfo{person}{Tobias Pape},
  \bibinfo{person}{Jeremy~G. Siek}, {and} \bibinfo{person}{Sam
  Tobin-Hochstadt}.} \bibinfo{year}{2015}\natexlab{}.
\newblock \showarticletitle{Pycket: A Tracing JIT for a Functional Language}.
  In \bibinfo{booktitle}{{\em Proceedings of the 20th ACM SIGPLAN International
  Conference on Functional Programming}} {\em (\bibinfo{series}{ICFP 2015})}.
  \bibinfo{publisher}{ACM}, \bibinfo{address}{New York, NY, USA},
  \bibinfo{pages}{22--34}.
\newblock
\showISBNx{978-1-4503-3669-7}
\showDOI{%
\url{http://dx.doi.org/10.1145/2784731.2784740}}


\bibitem[\protect\citeauthoryear{Bolz, Diekmann, and Tratt}{Bolz
  et~al\mbox{.}}{2013}]%
        {pypy}
\bibfield{author}{\bibinfo{person}{Carl~Friedrich Bolz}, \bibinfo{person}{Lukas
  Diekmann}, {and} \bibinfo{person}{Laurence Tratt}.}
  \bibinfo{year}{2013}\natexlab{}.
\newblock \showarticletitle{Storage Strategies for Collections in Dynamically
  Typed Languages}.
\newblock \bibinfo{journal}{{\em SIGPLAN Not.\/}} \bibinfo{volume}{48},
  \bibinfo{number}{10} (\bibinfo{date}{Oct.} \bibinfo{year}{2013}),
  \bibinfo{pages}{167--182}.
\newblock
\showISSN{0362-1340}
\showDOI{%
\url{http://dx.doi.org/10.1145/2544173.2509531}}


\bibitem[\protect\citeauthoryear{De~Wael, Marr, De~Koster, Sartor, and
  De~Meuter}{De~Wael et~al\mbox{.}}{2015}]%
        {DeWael:2015:JDS:2814228.2814231}
\bibfield{author}{\bibinfo{person}{Mattias De~Wael}, \bibinfo{person}{Stefan
  Marr}, \bibinfo{person}{Joeri De~Koster}, \bibinfo{person}{Jennifer~B.
  Sartor}, {and} \bibinfo{person}{Wolfgang De~Meuter}.}
  \bibinfo{year}{2015}\natexlab{}.
\newblock \showarticletitle{Just-in-time Data Structures}. In
  \bibinfo{booktitle}{{\em 2015 ACM International Symposium on New Ideas, New
  Paradigms, and Reflections on Programming and Software (Onward!)}} {\em
  (\bibinfo{series}{Onward! 2015})}. \bibinfo{publisher}{ACM},
  \bibinfo{address}{New York, NY, USA}, \bibinfo{pages}{61--75}.
\newblock
\showISBNx{978-1-4503-3688-8}
\showDOI{%
\url{http://dx.doi.org/10.1145/2814228.2814231}}


\bibitem[\protect\citeauthoryear{Gordon, Parkinson, Parsons, Bromfield, and
  Duffy}{Gordon et~al\mbox{.}}{2012}]%
        {Gordon:2012:URI:2384616.2384619}
\bibfield{author}{\bibinfo{person}{Colin~S. Gordon},
  \bibinfo{person}{Matthew~J. Parkinson}, \bibinfo{person}{Jared Parsons},
  \bibinfo{person}{Aleks Bromfield}, {and} \bibinfo{person}{Joe Duffy}.}
  \bibinfo{year}{2012}\natexlab{}.
\newblock \showarticletitle{Uniqueness and Reference Immutability for Safe
  Parallelism}.
\newblock \bibinfo{journal}{{\em SIGPLAN Not.\/}} \bibinfo{volume}{47},
  \bibinfo{number}{10} (\bibinfo{date}{Oct.} \bibinfo{year}{2012}),
  \bibinfo{pages}{21--40}.
\newblock
\showISSN{0362-1340}
\showDOI{%
\url{http://dx.doi.org/10.1145/2398857.2384619}}


\bibitem[\protect\citeauthoryear{Gramoli}{Gramoli}{2015}]%
        {Gramoli:2015:MYE:2688500.2688501}
\bibfield{author}{\bibinfo{person}{Vincent Gramoli}.}
  \bibinfo{year}{2015}\natexlab{}.
\newblock \showarticletitle{More Than You Ever Wanted to Know About
  Synchronization: Synchrobench, Measuring the Impact of the Synchronization on
  Concurrent Algorithms}. In \bibinfo{booktitle}{{\em Proceedings of the 20th
  ACM SIGPLAN Symposium on Principles and Practice of Parallel Programming}}
  {\em (\bibinfo{series}{PPoPP 2015})}. \bibinfo{publisher}{ACM},
  \bibinfo{address}{New York, NY, USA}, \bibinfo{pages}{1--10}.
\newblock
\showISBNx{978-1-4503-3205-7}
\showDOI{%
\url{http://dx.doi.org/10.1145/2688500.2688501}}


\bibitem[\protect\citeauthoryear{Harris}{Harris}{2001}]%
        {lazy-lists-Harris:2001}
\bibfield{author}{\bibinfo{person}{Timothy~L. Harris}.}
  \bibinfo{year}{2001}\natexlab{}.
\newblock \showarticletitle{A Pragmatic Implementation of Non-blocking
  Linked-Lists}. In \bibinfo{booktitle}{{\em Proceedings of the 15th
  International Conference on Distributed Computing}} {\em
  (\bibinfo{series}{DISC '01})}. \bibinfo{publisher}{Springer-Verlag},
  \bibinfo{address}{London, UK, UK}, \bibinfo{pages}{300--314}.
\newblock
\showISBNx{3-540-42605-1}
\showURL{%
\url{http://dl.acm.org/citation.cfm?id=645958.676105}}


\bibitem[\protect\citeauthoryear{Harris, Fraser, and Pratt}{Harris
  et~al\mbox{.}}{2002}]%
        {Harris:2002:PMC:645959.676137}
\bibfield{author}{\bibinfo{person}{Timothy~L. Harris}, \bibinfo{person}{Keir
  Fraser}, {and} \bibinfo{person}{Ian~A. Pratt}.}
  \bibinfo{year}{2002}\natexlab{}.
\newblock \showarticletitle{A Practical Multi-word Compare-and-Swap Operation}.
  In \bibinfo{booktitle}{{\em Proceedings of the 16th International Conference
  on Distributed Computing}} {\em (\bibinfo{series}{DISC '02})}.
  \bibinfo{publisher}{Springer-Verlag}, \bibinfo{address}{London, UK, UK},
  \bibinfo{pages}{265--279}.
\newblock
\showISBNx{3-540-00073-9}
\showURL{%
\url{http://dl.acm.org/citation.cfm?id=645959.676137}}


\bibitem[\protect\citeauthoryear{Hendler, Shavit, and Yerushalmi}{Hendler
  et~al\mbox{.}}{2004}]%
        {hendler-stack-2004}
\bibfield{author}{\bibinfo{person}{Danny Hendler}, \bibinfo{person}{Nir
  Shavit}, {and} \bibinfo{person}{Lena Yerushalmi}.}
  \bibinfo{year}{2004}\natexlab{}.
\newblock \showarticletitle{A Scalable Lock-free Stack Algorithm}. In
  \bibinfo{booktitle}{{\em Proceedings of the Sixteenth Annual ACM Symposium on
  Parallelism in Algorithms and Architectures}} {\em (\bibinfo{series}{SPAA
  '04})}. \bibinfo{publisher}{ACM}, \bibinfo{address}{New York, NY, USA},
  \bibinfo{pages}{206--215}.
\newblock
\showISBNx{1-58113-840-7}
\showDOI{%
\url{http://dx.doi.org/10.1145/1007912.1007944}}


\bibitem[\protect\citeauthoryear{Herlihy, Luchangco, and Moir}{Herlihy
  et~al\mbox{.}}{2003a}]%
        {Herlihy:2003:OSD:850929.851942}
\bibfield{author}{\bibinfo{person}{Maurice Herlihy}, \bibinfo{person}{Victor
  Luchangco}, {and} \bibinfo{person}{Mark Moir}.}
  \bibinfo{year}{2003}\natexlab{a}.
\newblock \showarticletitle{Obstruction-Free Synchronization: Double-Ended
  Queues As an Example}. In \bibinfo{booktitle}{{\em Proceedings of the 23rd
  International Conference on Distributed Computing Systems}} {\em
  (\bibinfo{series}{ICDCS '03})}. \bibinfo{publisher}{IEEE Computer Society},
  \bibinfo{address}{Washington, DC, USA}, \bibinfo{pages}{522--}.
\newblock
\showISBNx{0-7695-1920-2}
\showURL{%
\url{http://dl.acm.org/citation.cfm?id=850929.851942}}


\bibitem[\protect\citeauthoryear{Herlihy, Luchangco, Moir, and Scherer}{Herlihy
  et~al\mbox{.}}{2003b}]%
        {Herlihy:2003:STM:872035.872048}
\bibfield{author}{\bibinfo{person}{Maurice Herlihy}, \bibinfo{person}{Victor
  Luchangco}, \bibinfo{person}{Mark Moir}, {and} \bibinfo{person}{William~N.
  Scherer, III}.} \bibinfo{year}{2003}\natexlab{b}.
\newblock \showarticletitle{Software Transactional Memory for Dynamic-sized
  Data Structures}. In \bibinfo{booktitle}{{\em Proceedings of the
  Twenty-second Annual Symposium on Principles of Distributed Computing}} {\em
  (\bibinfo{series}{PODC '03})}. \bibinfo{publisher}{ACM},
  \bibinfo{address}{New York, NY, USA}, \bibinfo{pages}{92--101}.
\newblock
\showISBNx{1-58113-708-7}
\showDOI{%
\url{http://dx.doi.org/10.1145/872035.872048}}


\bibitem[\protect\citeauthoryear{Herlihy and Wing}{Herlihy and Wing}{1990}]%
        {herlihy1990linearizability}
\bibfield{author}{\bibinfo{person}{Maurice~P. Herlihy} {and}
  \bibinfo{person}{Jeannette~M. Wing}.} \bibinfo{year}{1990}\natexlab{}.
\newblock \showarticletitle{Linearizability: A Correctness Condition for
  Concurrent Objects}.
\newblock \bibinfo{journal}{{\em ACM Trans. Program. Lang. Syst.\/}}
  \bibinfo{volume}{12}, \bibinfo{number}{3} (\bibinfo{date}{July}
  \bibinfo{year}{1990}), \bibinfo{pages}{463--492}.
\newblock
\showISSN{0164-0925}
\showDOI{%
\url{http://dx.doi.org/10.1145/78969.78972}}


\bibitem[\protect\citeauthoryear{Kilpatrick, Dreyer, Peyton~Jones, and
  Marlow}{Kilpatrick et~al\mbox{.}}{2014}]%
        {kilpatrick2014backpack}
\bibfield{author}{\bibinfo{person}{Scott Kilpatrick}, \bibinfo{person}{Derek
  Dreyer}, \bibinfo{person}{Simon Peyton~Jones}, {and} \bibinfo{person}{Simon
  Marlow}.} \bibinfo{year}{2014}\natexlab{}.
\newblock \showarticletitle{Backpack: retrofitting Haskell with interfaces}. In
  \bibinfo{booktitle}{{\em ACM SIGPLAN Notices}}, Vol.~\bibinfo{volume}{49}.
  ACM, \bibinfo{pages}{19--31}.
\newblock


\bibitem[\protect\citeauthoryear{Kusum, Neamtiu, and Gupta}{Kusum
  et~al\mbox{.}}{2016}]%
        {Kusum:2016:SFA:2892208.2892220}
\bibfield{author}{\bibinfo{person}{Amlan Kusum}, \bibinfo{person}{Iulian
  Neamtiu}, {and} \bibinfo{person}{Rajiv Gupta}.}
  \bibinfo{year}{2016}\natexlab{}.
\newblock \showarticletitle{Safe and Flexible Adaptation via Alternate Data
  Structure Representations}. In \bibinfo{booktitle}{{\em Proceedings of the
  25th International Conference on Compiler Construction}} {\em
  (\bibinfo{series}{CC 2016})}. \bibinfo{publisher}{ACM}, \bibinfo{address}{New
  York, NY, USA}, \bibinfo{pages}{34--44}.
\newblock
\showISBNx{978-1-4503-4241-4}
\showDOI{%
\url{http://dx.doi.org/10.1145/2892208.2892220}}


\bibitem[\protect\citeauthoryear{Leino, M\"{u}ller, and Wallenburg}{Leino
  et~al\mbox{.}}{2008}]%
        {Leino:2008:FIF:1434628.1434650}
\bibfield{author}{\bibinfo{person}{K.~Rustan Leino}, \bibinfo{person}{Peter
  M\"{u}ller}, {and} \bibinfo{person}{Angela Wallenburg}.}
  \bibinfo{year}{2008}\natexlab{}.
\newblock \showarticletitle{Flexible Immutability with Frozen Objects}. In
  \bibinfo{booktitle}{{\em Proceedings of the 2Nd International Conference on
  Verified Software: Theories, Tools, Experiments}} {\em
  (\bibinfo{series}{VSTTE '08})}. \bibinfo{publisher}{Springer-Verlag},
  \bibinfo{address}{Berlin, Heidelberg}, \bibinfo{pages}{192--208}.
\newblock
\showISBNx{978-3-540-87872-8}
\showDOI{%
\url{http://dx.doi.org/10.1007/978-3-540-87873-5_17}}


\bibitem[\protect\citeauthoryear{Marlow, Newton, and Peyton~Jones}{Marlow
  et~al\mbox{.}}{2011}]%
        {monad-par}
\bibfield{author}{\bibinfo{person}{Simon Marlow}, \bibinfo{person}{Ryan~R.
  Newton}, {and} \bibinfo{person}{Simon Peyton~Jones}.}
  \bibinfo{year}{2011}\natexlab{}.
\newblock \showarticletitle{A monad for deterministic parallelism}. In
  \bibinfo{booktitle}{{\em Proceedings of the 4th ACM symposium on {H}askell}}
  {\em (\bibinfo{series}{Haskell '11})}. \bibinfo{publisher}{ACM},
  \bibinfo{pages}{71--82}.
\newblock


\bibitem[\protect\citeauthoryear{Nawab, Chakrabarti, Kelly, and
  Morrey~III}{Nawab et~al\mbox{.}}{2015}]%
        {lock-free-crash-resilience}
\bibfield{author}{\bibinfo{person}{Faisal Nawab}, \bibinfo{person}{Dhruva~R
  Chakrabarti}, \bibinfo{person}{Terence Kelly}, {and}
  \bibinfo{person}{Charles~B Morrey~III}.} \bibinfo{year}{2015}\natexlab{}.
\newblock \showarticletitle{Procrastination Beats Prevention: Timely Sufficient
  Persistence for Efficient Crash Resilience.}. In \bibinfo{booktitle}{{\em
  EDBT}}. \bibinfo{pages}{689--694}.
\newblock


\bibitem[\protect\citeauthoryear{Newton, Fogg, and Varamesh}{Newton
  et~al\mbox{.}}{2015}]%
        {adaptive-data-icfp15}
\bibfield{author}{\bibinfo{person}{Ryan~R. Newton}, \bibinfo{person}{Peter~P.
  Fogg}, {and} \bibinfo{person}{Ali Varamesh}.}
  \bibinfo{year}{2015}\natexlab{}.
\newblock \showarticletitle{Adaptive Lock-free Maps: Purely-functional to
  Scalable}. In \bibinfo{booktitle}{{\em Proceedings of the 20th ACM SIGPLAN
  International Conference on Functional Programming}} {\em
  (\bibinfo{series}{ICFP 2015})}. \bibinfo{publisher}{ACM},
  \bibinfo{address}{New York, NY, USA}, \bibinfo{pages}{218--229}.
\newblock
\showISBNx{978-1-4503-3669-7}
\showDOI{%
\url{http://dx.doi.org/10.1145/2784731.2784734}}


\bibitem[\protect\citeauthoryear{Nikhil}{Nikhil}{1991}]%
        {id}
\bibfield{author}{\bibinfo{person}{Rishiyur~S. Nikhil}.}
  \bibinfo{year}{1991}\natexlab{}.
\newblock \bibinfo{title}{ID Language Reference Manual}.
\newblock   (\bibinfo{year}{1991}).
\newblock


\bibitem[\protect\citeauthoryear{Peyton~Jones, Gordon, and Finne}{Peyton~Jones
  et~al\mbox{.}}{1996}]%
        {concurrent-haskell-popl96}
\bibfield{author}{\bibinfo{person}{Simon Peyton~Jones}, \bibinfo{person}{Andrew
  Gordon}, {and} \bibinfo{person}{Sigbjorn Finne}.}
  \bibinfo{year}{1996}\natexlab{}.
\newblock \showarticletitle{Concurrent Haskell}. In \bibinfo{booktitle}{{\em
  Proceedings of the 23rd ACM SIGPLAN-SIGACT Symposium on Principles of
  Programming Languages}} {\em (\bibinfo{series}{POPL '96})}.
  \bibinfo{publisher}{ACM}, \bibinfo{address}{New York, NY, USA},
  \bibinfo{pages}{295--308}.
\newblock
\showISBNx{0-89791-769-3}
\showDOI{%
\url{http://dx.doi.org/10.1145/237721.237794}}


\bibitem[\protect\citeauthoryear{Peyton~Jones, Reid, Henderson, Hoare, and
  Marlow}{Peyton~Jones et~al\mbox{.}}{1999}]%
        {haskell-imprecise-exceptions}
\bibfield{author}{\bibinfo{person}{Simon Peyton~Jones},
  \bibinfo{person}{Alastair Reid}, \bibinfo{person}{Fergus Henderson},
  \bibinfo{person}{Tony Hoare}, {and} \bibinfo{person}{Simon Marlow}.}
  \bibinfo{year}{1999}\natexlab{}.
\newblock \showarticletitle{A semantics for imprecise exceptions}. In
  \bibinfo{booktitle}{{\em ACM SIGPLAN Notices}}, Vol.~\bibinfo{volume}{34}.
  ACM, \bibinfo{pages}{25--36}.
\newblock


\bibitem[\protect\citeauthoryear{Prokopec, Bronson, Bagwell, and
  Odersky}{Prokopec et~al\mbox{.}}{2012}]%
        {Prokopec:2012:CTE:2145816.2145836}
\bibfield{author}{\bibinfo{person}{Aleksandar Prokopec},
  \bibinfo{person}{Nathan~Grasso Bronson}, \bibinfo{person}{Phil Bagwell},
  {and} \bibinfo{person}{Martin Odersky}.} \bibinfo{year}{2012}\natexlab{}.
\newblock \showarticletitle{Concurrent Tries with Efficient Non-blocking
  Snapshots}. In \bibinfo{booktitle}{{\em Proceedings of the 17th ACM SIGPLAN
  Symposium on Principles and Practice of Parallel Programming}} {\em
  (\bibinfo{series}{PPoPP '12})}. \bibinfo{publisher}{ACM},
  \bibinfo{address}{New York, NY, USA}, \bibinfo{pages}{151--160}.
\newblock
\showISBNx{978-1-4503-1160-1}
\showDOI{%
\url{http://dx.doi.org/10.1145/2145816.2145836}}


\bibitem[\protect\citeauthoryear{Shalev and Shavit}{Shalev and Shavit}{2006}]%
        {split-ordered-list}
\bibfield{author}{\bibinfo{person}{Ori Shalev} {and} \bibinfo{person}{Nir
  Shavit}.} \bibinfo{year}{2006}\natexlab{}.
\newblock \showarticletitle{Split-ordered Lists: Lock-free Extensible Hash
  Tables}.
\newblock \bibinfo{journal}{{\em J. ACM\/}} \bibinfo{volume}{53},
  \bibinfo{number}{3} (\bibinfo{date}{May} \bibinfo{year}{2006}),
  \bibinfo{pages}{379--405}.
\newblock
\showISSN{0004-5411}
\showDOI{%
\url{http://dx.doi.org/10.1145/1147954.1147958}}


\bibitem[\protect\citeauthoryear{Vollmer, Scott, Musuvathi, and Newton}{Vollmer
  et~al\mbox{.}}{2017}]%
        {sc-haskell-draft}
\bibfield{author}{\bibinfo{person}{Michael Vollmer}, \bibinfo{person}{Ryan~G.
  Scott}, \bibinfo{person}{Madan Musuvathi}, {and} \bibinfo{person}{Ryan~R.
  Newton}.} \bibinfo{year}{2017}\natexlab{}.
\newblock \showarticletitle{SC-Haskell: Sequential Consistency in Languages
  that Minimize Mutable Shared Heap}. In \bibinfo{booktitle}{{\em To appear in
  the proceedings of the 22st ACM SIGPLAN Symposium on Principles and Practice
  of Parallel Programming}} {\em (\bibinfo{series}{PPoPP '17})}.
  \bibinfo{publisher}{ACM}, \bibinfo{address}{New York, NY, USA}.
\newblock


\bibitem[\protect\citeauthoryear{Xu}{Xu}{2013}]%
        {java-collection-swapper}
\bibfield{author}{\bibinfo{person}{Guoqing Xu}.}
  \bibinfo{year}{2013}\natexlab{}.
\newblock \showarticletitle{CoCo: Sound and Adaptive Replacement of Java
  Collections}. In \bibinfo{booktitle}{{\em Proceedings of the 27th European
  Conference on Object-Oriented Programming}} {\em
  (\bibinfo{series}{ECOOP'13})}. \bibinfo{publisher}{Springer-Verlag},
  \bibinfo{address}{Berlin, Heidelberg}, \bibinfo{pages}{1--26}.
\newblock
\showISBNx{978-3-642-39037-1}
\showDOI{%
\url{http://dx.doi.org/10.1007/978-3-642-39038-8_1}}


\bibitem[\protect\citeauthoryear{Yang, Campagna, A\u{g}acan, El-Hassany,
  Kulkarni, and Newton}{Yang et~al\mbox{.}}{2015}]%
        {cnf-icfp15}
\bibfield{author}{\bibinfo{person}{Edward~Z. Yang}, \bibinfo{person}{Giovanni
  Campagna}, \bibinfo{person}{\"{O}mer~S. A\u{g}acan}, \bibinfo{person}{Ahmed
  El-Hassany}, \bibinfo{person}{Abhishek Kulkarni}, {and}
  \bibinfo{person}{Ryan~R. Newton}.} \bibinfo{year}{2015}\natexlab{}.
\newblock \showarticletitle{Efficient Communication and Collection with Compact
  Normal Forms}. In \bibinfo{booktitle}{{\em Proceedings of the 20th ACM
  SIGPLAN International Conference on Functional Programming}} {\em
  (\bibinfo{series}{ICFP 2015})}. \bibinfo{publisher}{ACM},
  \bibinfo{address}{New York, NY, USA}, \bibinfo{pages}{362--374}.
\newblock
\showISBNx{978-1-4503-3669-7}
\showDOI{%
\url{http://dx.doi.org/10.1145/2784731.2784735}}


\end{thebibliography}
}


\end{document}